\documentclass[twocolumn,amsmath,amssymb,prb,superscriptaddress,a4paper,floatfix,showkeys]{revtex4}
\usepackage{dcolumn}
\usepackage{graphicx}
\usepackage{epsfig}

\newcommand\sumprime{\mathop{{\sum}'}}
\newcommand\MT{Maxwell's theory}
\newcommand\QT{quantum theory}

\begin{document}

\title{Event-based Corpuscular Model for Quantum Optics Experiments}

\author{Kristel Michielsen}
\email{k.michielsen@fz-juelich.de}           
\affiliation{%
Institute for Advanced Simulation, J\"ulich Supercomputing Centre,
Research Centre J\"ulich, D-52425 J\"ulich, Germany
}%
\author{F. Jin}
\email{f.jin@rug.nl}
\affiliation{%
Department of Applied Physics, Zernike Institute for Advanced Materials,
University of Groningen, Nijenborgh 4, NL-9747 AG Groningen, The Netherlands
}%
\author{H. De Raedt}
\email{h.a.de.raedt@rug.nl}
\affiliation{%
Department of Applied Physics, Zernike Institute for Advanced Materials,
University of Groningen, Nijenborgh 4, NL-9747 AG Groningen, The Netherlands
}%


\date{\today}

\begin{abstract}
A corpuscular simulation model of optical phenomena that does not require the knowledge of the solution of
a wave equation of the whole system and reproduces the results of \MT\ by generating detection events one-by-one is presented.
The event-based corpuscular model is shown to give a unified description of multiple-beam fringes of a plane parallel plate,
single-photon Mach-Zehnder interferometer, Wheeler's delayed choice, photon tunneling,
quantum erasers, two-beam interference, double-slit,
and Einstein-Podolsky-Rosen-Bohm and Hanbury Brown-Twiss experiments. 
\end{abstract}

\maketitle
\tableofcontents

\section{Introduction}

The theory of optical phenomena has a long and interesting history~\cite{BORN64,PAUL04,GARR09}.
The corpuscular theory of Newton and his followers
was abandoned in favor of extensions pf Huygens' wave theory,
culminating in \MT\ of electrodynamics.
\MT\ is extremely powerfull.
It applies to virtually all electrodynamic phenomena that find practical, real-life applications.
Yet, as with any theory, \MT\ has its limitations.
With Einstein's explanation of the photoelectric effect in terms of photons,
that is in terms of indivisible quanta of light,
the idea of a corpuscular description of light revived.
Einstein's hypothesis of light quanta gave birth to the quantum description of light.
As the photoelectric effect can be explained
by treating the electromagnetic field without assuming the existence
of photons~\cite{GARR09}, the photoelectric effect itself
does not indicate that light consists of indivisible particles
but the experiments to be described next do.

\subsection{Photon indivisability experiments}\label{pie}

The paper by Grangier {\sl et al.}~\cite{GRAN86} reports clear and direct evidence for the indivisiblity of the
single photons~\cite{GARR09}.
A key feature in this work is the use of the three-level cascade photon emission of the calcium atom.
When the calcium atoms are excited to the third lowest level, they relax to the second lowest state, emitting photons of frequency
$f$, followed by another transition to the ground state level causing photons of frequency $f'$ to be emitted~\cite{KOCH67}.
It is observed that each such two-step process emits two photons in two spatially well-separated directions,
allowing for the cascade emission to be detected using a time-coincidence technique~\cite{KOCH67}.

\begin{figure}[t]
\begin{center}
\includegraphics[width=7.5cm]{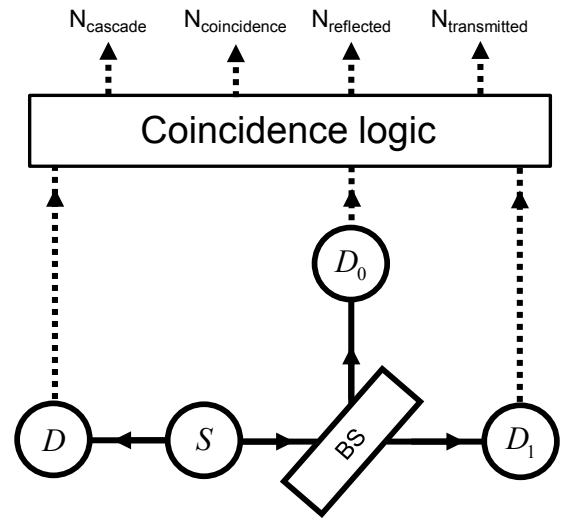}
\caption{Schematic diagram of the first experiment reported in
the paper by Grangier {\sl et al.}~\cite{GRAN86}, demonstrating the indivisiblity of the photon.
S: Light source;
BS: 50-50 beam splitter;
$D$, $D_0$, $D_1$: Detectors
$N_{\mathbf{cascade}}$: Number of excitations of the three-level cascade of the calcium atoms in the source S;
$N_{\mathbf{coincidence}}$: Number of times detector $D_0$ and $D_1$ fire within a time window relative to the excitation of the cascade~\cite{GRAN86};
$N_{\mathbf{reflectied}}$: Number of photons reflected by the BS;
$N_{\mathbf{transmitted}}$: Number of photons transmitted by the BS.
Solid lines: Flow of photons.
Dashed lines: Flow of data.
}
\label{figgrang1}
\end{center}

\end{figure}

In Fig.~\ref{figgrang1} we show a diagram of the first experiment reported in Ref.\onlinecite{GRAN86}.
One of two light beams produced by the cascade is directed to detector $D$.
The other beam is sent through a 50-50 beam splitter to detectors $D_0$ and $D_1$.
Time-coincindence logic is used to establish the emission of the photons by the three level cascade process:
Only if detector $D$ and $D_0$, $D_1$ or both fire, a cascade emission event occured.
Then, the absence of a coincidence between the firing of detectors $D_0$ and $D_1$
provides unambigous evidence that the photon created in the cascade and passing through the beam splitter
behaves as one indivisble entity.
The analysis of the experimental data strongly support the hypothesis that
the photons created by the cascade process in the calcium atom are to be regarded as indivisible~\cite{GARR09}.

\begin{figure}[t]
\begin{center}
\includegraphics[width=8.5cm]{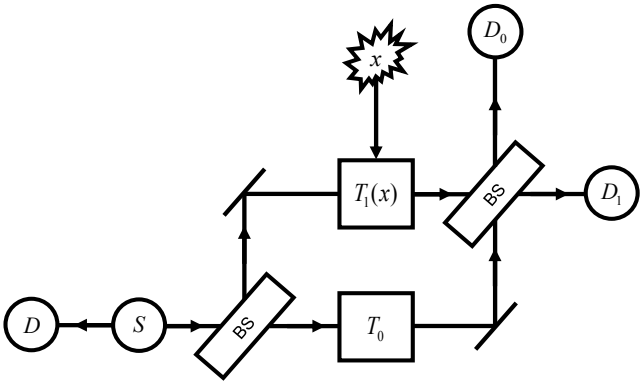}
\caption{Schematic diagram of the second experiment reported in
the paper by Grangier {\sl et al.}~\cite{GRAN86}, demonstrating that photons build up
an interference pattern one by one.
S: Light source;
BS: 50-50 beam splitter;
$T_0$: Fixed time-of-flight;
$T_1(x)$: Variable time-of-flight controlled by the external variable $x$;
$D_{0}$, $D_{1}$: Detectors.
For clarity, the coincidence logic (see Fig.~\ref{figgrang2}) has been omitted.}
\label{figgrang2}
\label{fig1}
\end{center}
\end{figure}

Having established the corpuscular nature of single photons,
Grangier {\sl et al.}~\cite{GRAN86} extend the experiment shown in Fig.~\ref{figgrang1}
by sending the photons emerging from the beam splitter to another beam splitter,
as illustrated in Fig.~\ref{figgrang2}, thereby constructing a Mach-Zehnder interferometer (MZI).
As the mirrors $M$ do not alter the particle character of the photons,
the removal of the second beam splitter yields an
experimental configuration that is equivalent to the one used to demonstrate
the corpuscular nature of single photons.
With the second beam splitter in place, Grangier {\sl et al.}~\cite{GRAN86}
observe that after collecting many photons one-by-one, the detection counts
fit nicely to the interference curve that is predicted by \MT,
that is what they observe is the same result as if the source would have emitted a wave.
Thus, it is shown that one-by-one these particles can build up an interference pattern,
just as in the experiment with single-electrons~\cite{MERL76,TONO89,TONO98}, for instance.


\subsection{Particles, waves or magic?}

In summary, the experimental observations lead to the conclusions that
\begin{enumerate}
\item{Individual photons are indivisible}
\item{In a MZI, the collection of many individual photons produces the interference pattern that is expected from classical wave theory.}
\end{enumerate}

The conjuction of these two conclusions cannot be explained within \MT\ or \QT.
This is most obvious in the case of \MT\ which does not pretend to describe particles.
Imagining a single photon to be a spatially localized excitation of a wave field,
this so-called wave packet would, according to wave theory, be divided into two wave packet by a beam splitter.
This is not what is observed in the experiments of Grangier {\sl et al.}~\cite{GRAN86}.

Particle-wave duality, a concept of \QT\
which attributes to photons the properties of both wave and particle depending upon the circumstances,
does not help to explain the experimental facts either.
The experiments~\cite{GRAN86} show that each individual photon behaves as a particle, not as a wave.
In these experiments (and in many others that use single-photon sources),
it is clear that one particular photon never interferes with itself nor with other ones;
the wave functions that are used in the wave mechanical theory interfere if
they interact with material only~\cite{ROYC10}.
It is only in the mathematical, statistical description of many detected photons
that (probability) waves interfere~\cite{MAND99}.

As a last resort to explain the experimental facts using concepts of \QT\ the idea of wavefunction collapse~\cite{NEUM55} is often used.
According to this idea, the wavefunction materializes into a particle during the act of measurement.
The mechanism that gives rise to this collapse has remained elusive for 78 years after of its conception.
Therefore, with the present state of understanding, invoking the wavefunction collapse
to explain an experimental observation adds a flavor of mysticism to the explanation.
It is important to note that the mystical element never enters
a \QT\ calculation of the statistical averages~\cite{BALL03} and is,
for any practical purpose, superfluous.
It only serves to create the illusion that \QT\ has something
meaningfull to say about an individual event but this is not the case.
Quantum theory gives us a recipe to compute the frequencies (averages) for observing events
but does not describe individual event themselves~\cite{HOME97}.
Of course, the introduction of a mystical element is undesirable from a scientific viewpoint and,
as shown in the present paper, also unnecessary to explain the observed phenomena.

\subsection{Aim of this work}

The present work is based on three assumptions, namely
\begin{enumerate}
\item{A photon is an indivisible entity}.
\item{Photons do not interact}.
\item{The result of many photons recorded by a detector is described by \MT}.
\end{enumerate}
Obviously, there is strong emperical evidence for all these three assumptions.
As explained earlier, the conjunction of assumption 1 and 3 poses some serious conceptual problems
that cannot be resolved within \QT\ proper.

In this paper, we show that quantum optics experiments which are performed in the single-photon regime (as in the experiments
discussed in Section~\ref{pie}), can be explained entirely
\begin{itemize}
\item{with a universal event-based corpuscular model},
\item{without first solving a wave equation},
\end{itemize}
The event-based corpuscular model (EBCM) that we introduce in this paper is universal in that it can, without modification,
be used to explain why photons build up interference patterns, why they can exhibit correlations that
cannot be explained within \MT\ and so on.

The EBCM gives a cause-and-effect description for every step of the process,
starting with the emission and ending with the detection of the photon.
By construction, the EBCM satisfies Einstein's criterion of local causality.
Although not essential, an appealing feauture of the EBCM is that it allows for a realistic interpretation
of the variables that appear in the simulation model.
The research presented in this paper unifies and extends
earlier work~\cite{RAED05d,RAED05b,RAED05c,MICH05,RAED06c,RAED07a,RAED07b,RAED07c,ZHAO07b,ZHAO08,ZHAO08b,JIN09a,JIN09c,JIN10a}.

\subsection{Disclaimer}

The present paper is not concerned with an interpretation or an extension of \QT.
The existence of an EBCM for event-based phenomena does not affect
the validity and applicability of \QT\ or \MT\ but shows
that it is possible to give explanations of observed phenomena that
do not find a logically consistent, rational explanation within these two theories.

\subsection {Structure of the paper}
The paper is organized as follows.
In Section~\ref{comput}, we examine the question of particle-versus-wave
from a computational point of view.
We show that also from this viewpoint, there are conceptual difficulties
to merge the wave description in terms of probability amplitudes
with the fact that detectors ``click".
Section~\ref{sec3} discusses the general ideas that underpin our event-by-event simulation approach.
In Section~\ref{sec5} we introduce the EBCM of the interface between two media and of the detector,
that is we define the algorithms that will be used in the particle-by-particle simulation.
Section~\ref{validation} is devoted to the validation of the EBCM:
It is shown that it reproduces the results of \MT\ for a simple
interface, a plane-parallel plate, and the double-slit experiment,
without changing the algorithms of course.
Building on the EBCM of the interface, Section~\ref{sec5} specifies the EBCM of the optical components that
are used in (quantum) optics.
In Section~\ref{sec6}, we show that the same EBCM reproduces,
event-by-event and without using the solution of a wave equation,
Mach-Zehnder interferometer experiments~\cite{GRAN86},
Wheeler's delayed choice experiment~\cite{JACQ07},
quantum eraser~\cite{SCHW99} and photon tunneling experiments~\cite{MIZO92,UNNI96,BRID04}.
Section~\ref{PCE} shows that the EBCM approach effortlessly extends
to experiments with correlated photons
by showing that it reproduces the \QT\ results
of the Einstein-Podolsky-Rosen-Bohm (EPRB) experiments with photons~\cite{ASPE82a,ASPE82b,WEIH98}
and Hanbury Brown-Twiss (HBT) experiments.
Our conclusions and a discussion of open problems are given in Section~\ref{summary}.

\section{Computational point of view}\label{comput}

Our approach employs algorithms, that is we define processes, that contain
a detailed specification of each individual event
which, as we now show by an explicit example,
cannot be derived from a wave theory such as Maxwell's theory or \QT.

To understand the subtleties that are involved, it is helpful to consider
as a simple example, the conventional wave theoretical description
of the Mach-Zehnder interferometer.

\subsubsection{Mach-Zehnder interferometer}

A schematic diagram of a MZI experiment is shown in Fig.~\ref{fig1}.
We assume that the length $L_0=cT_0$ of the lower arm of the interferometer
is fixed and that the length $L_1(x)=cT_1(x)$ of the upper arm can be varied by changing the control variable $x$.

Assuming a coherent monochromatic light source $S$ with frequency $\omega$
and a fixed value $x$, it follows from Maxwell's theory~\cite{BORN64} that the electric-field
amplitudes $b_0$ and $b_1$ on detectors 0 and 1 are related to the input amplitudes
$a_0$ and $a_1$ by
\begin{eqnarray}
\left(
\begin{array}{c}
b_0\\
b_1
\end{array}
\right)
&=&
\frac{1}{2}
\left(
\begin{array}{cc}
1&i\\
i&1
\end{array}
\right)
\left(
\begin{array}{cc}
e^{i\phi_0}&0\\
0&e^{i\phi_1}
\end{array}
\right)
\left(
\begin{array}{cc}
1&i\\
i&1
\end{array}
\right)
\left(
\begin{array}{c}
a_0\\
a_1
\end{array}
\right)
,\cr
&\equiv & ABA
\left(
\begin{array}{c}
a_0\\
a_1
\end{array}
\right)
,
\label{MZ1}
\end{eqnarray}
where
\begin{eqnarray}
A=\frac{1}{\sqrt{2}}
\left(
\begin{array}{cc}
1&i\\
i&1
\end{array}
\right)
\quad,\quad
B=
\left(
\begin{array}{cc}
e^{i\phi_0}&0\\
0&e^{i\phi_1}
\end{array}
\right)
,
\label{MZ1a}
\end{eqnarray}
where $\phi_0=\omega T_0$ and $\phi_1(x)=\omega T_1(x)$.
As the light enters the MZI via port 0 of the left most
BS, we have $a_0=1$ and $a_1=0$ up to an irrelevant phase factor.
Then, according to \QT, the probabilities $P_0$ ($P_1$) for detectors $D_0$ or (exclusive) $D_1$ to generate a click,
are given by
\begin{equation}
P_k
=
\left|
\sum_{j=0,1}
\sum_{i=0,1}
A_{k,j}B_{j,i}A_{i,0}
\right|^2
\quad,\quad k=0,1
,
\label{MZ2}
\end{equation}
respectively.

Using Eqs.~(\ref{MZ1}) and (\ref{MZ2}) a simple calculation yields a closed form
expression for $P_k$, namely
\begin{eqnarray}
P_{0}&=&\sin^2\frac{\omega(T_0-T_1(x))}{2}=\sin^2\frac{\phi_0-\phi_1(x)}{2}
,
\label{mzi0a}
\\
P_{1}&=&\cos^2\frac{\omega(T_0-T_1(x))}{2}=\cos^2\frac{\phi_0-\phi_1(x)}{2}
.
\label{mzi0b}
\end{eqnarray}

Equations~(\ref{mzi0a}) and (\ref{mzi0b}) show that the signal on the detectors is modulated by the difference between
the time-of-flights $T_0$ and $T_1(x)$ in the lower and upper arm
of the interferometer, respectively, or in other words by the phase difference $\phi_0-\phi_1(x)$.
This modulation is characteristic for interference phenomena.

\subsubsection{Using the solution of wave theory}
Once we know $P_k$, it is trivial to construct a process that generates clicks of the detectors $D_0$ and $D_1$.
For instance, if we want to mimic the unpredictable character of the
single-photon detection process, we may use a pseudo-random number generator
to produce detection events.
It is irrelevant whether we have a closed form expression for $P_k$ or only know
$P_k$ in tabulated form. The key point is that we worked out the
sums over the indices $i$ and $j$ in Eq.~(\ref{MZ2}) ourselves,
and that the solution of the full wave mechanical problem
is used to produce the events.

\subsubsection{Fundamental problem: Example}

Let us now assume that we do not know how to perform the
sums over the indices $i$ and $j$ in Eq.~(\ref{MZ2}).
In other words, we assume that we do not know the explicit (or
tabulated) form of $P_0$ and $P_1$.
Then, we cannot simply use the pseudo-random number generator
to generate detector clicks.

Of course, we can still construct discrete-event processes that perform the sums
in Eq.~(\ref{MZ2}) by selecting (one-by-one) the pairs $(i,j)$ from the set
${\cal S}={(0,0),(1,0),(0,1),(1,1)}$.
Any such process defines a sequence of ``events'' $(i,j)$.
The key question now is: Can we identify the selection
of the pairs with ``clicks'' that correspond to detection events?
We now provide a trivial, rigorous proof that this is fundamentally impossible.

A characteristic feature of all wave phenomena
is that not all contributions to the sums
in Eq.~(\ref{MZ2}) have the same sign:
In wave theory, this feature is essential
to account for destructive interference.
But, at the same time this feature
forbids the existence of a process of which the ``events'' can
be identified with the clicks of the detector.

This is easily seen by considering a situation in which, for instance, $P_0=0$.
In this case, the detector $D_0$ should never click.
However, according to Eq.~(\ref{MZ2}), any process that samples
from the set ${\cal S}$ produces ``events''
such that the sum over all these ``events'' vanishes.
Therefore, if we want to identify these ``events''
with the clicks that we observe, we run into a logical contradiction:
To perform the sums in Eq.~(\ref{MZ2}), we have to generate events
that in the end cannot be interpreted as clicks since in this
particular case no detector clicks are observed.

\subsubsection{Fundamental problem: Generic case}

From the point of view of wave theory, the example of the MZI, though very simple,
is generic: Charateristic of Maxwell's theory and \QT\  is that the observable
phenomena are described by expressions, such as Eq.~(\ref{MZ2}),
that involve taking the square of the sum over amplitudes that are real or complex valued.
In general, these expressions take the form
\begin{equation}
P_{i_{n+1},i_{0}}
=
\left|
\sum_{\{i_1,\ldots,i_n\}}
A_{i_{n+1},i_n}^{(n)} \ldots A_{i_1,i_0}^{(1)}
\right|^2
,
\label{MZ3}
\end{equation}
where the elements of the matrices $A^{(j)}$
are, in general, complex valued, and the sums
over the indices $i_j$ can take discrete values or, as in the case
of a path integral, can be continuous variables.

As one or more of the $P_{i_{n+1},i_{0}}$'s may be zero,
the logically inescapable conclusion is
that the individual terms that appear in Eq.~(\ref{MZ3})
do not contain the necessary ingredients
to define a process that generates elements of the set $\{i_1,\ldots,i_n\}$
if each such element is to correspond to an observable (in experiment) event.

It is of interest to note that if the matrices
$A^{(j)}$ would have all non-negative entries,
each element of the set $\{i_1,\ldots,i_n\}$
would make a non-negative contribution to the total and hence,
it is possible to associate an event to each $n$-tuple $(i_1,\ldots,i_n)$.
Of course, in this case, the events can never form the interference patterns
that are characteristic of wave phenomena.

\subsubsection{A way out}

The crux of our event-by-event simulation approach is that we do not start
from an expression such as Eq.~(\ref{MZ3}) but construct a
classical, dynamical, event-by-event process that,
while generating events that correspond to the observed events,
produces these events with a frequency distributions that
converges to the unknown (by assumption) probability distribution
$P_{i_{n+1},i_{0}}$.

Initially, the system does not know about this limiting distribution and hence, during
a transient period, the frequencies with which events are generated
do not necessarily agree with this limiting distribution.
However, as ample numerical simulations demonstrate, for many events,
these first few ``wrong'' events do not significantly contribute to the averages
and are therefore irrelevant for the comparison of the event-based simulation results
with those of a wave theory.

\section{Event-by-event simulation}\label{sec3}

Our event-based simulation approach is unconventional in that it does not require
knowledge of the wave amplitudes obtained by first solving a wave mechanical problem.
Instead, the detector clicks are generated event-by-event by locally causal,
adaptive, classical dynamical systems.
In this section, we discuss the general aspects of our simulation approach.

The event-by-event simulation algorithms are most easily formulated in terms of messengers that
carry messages and units that accept messengers, process their messages and send the messengers with
a possibly modified message to the next unit.
A source is a simple unit that creates messengers with specified messages on demand.
A message consists of a set of numbers, conveniently represented by a vector.
A messenger that appears at the output of a detection unit corresponds to an event observed in experiment.
The processing units mimic the role of the optical components in the experiment
and the network by connecting the processing units represents the complete experimental setup.

Within the realm of quantum optics experiments, in a pictorial description, the photon is regarded as a messenger,
carrying a message that represents its time-of-flight (phase) and polarization.
In this pictorial description, we may speak of ``photons'' generating the detection events.
However, these so-called photons, as we will call them in the following,
are elements of a model or theory for the real laboratory experiment only.
The only experimental facts are the settings of the various apparatuses and the detection events.
What happens in between activating the source and the registration of the detection
events belongs to the domain of imagination.

In general, a processing unit consist of an input stage, a transformation stage and an output stage.
The input (output) stage may have several ports at (through) which messengers arrive (leave).
As a messenger arrives at an input port of a processing unit,
the input stage updates its internal state, also represented by a vector.
This update mechanism renders the unit adaptive and is essential for the simulation approach
to reproduce the results of wave mechanics.
For reasons that are explained later, this input stage is called deterministic learning machine (DLM).
Next, the message together with the internal state of the unit are passed to
the transformation stage that implements the operation of the particular device.
Finally, a new message is sent to the output stage
which selects the output port through which the messenger will leave the unit.
Some processing units are simpler in the sense that the input stage
may be omitted, in which case the unit lacks adaptivity.
The rule to decide which type of unit to employ is simple:
If the unit should be able to deal with ``interference phenomena'' ,
it should be adaptive, otherwise we may (but do not have to) use
the simpler device. Interfaces between materials of different index of refraction,
beams splitters and the like belong to the first class. A device that
causes a time delay only belongs to the second class.

A very important feature of our simulation approach is that
at any given time, there is only one messenger being routed through the whole network.
Therefore there can be no direct communication between the messengers.
As the internal state of a unit may depend on the messages it has already
processed, a unit can modify the messenger and its message.
It is this mechanism that can cause ``interference without waves''.
From this general description of the simulation algorithm,
it should already be clear that the process that
is being generated by the network of processing units
is locally causal in Einstein's sense.

The processing units that are essential for our approach
to reproduce the results of a wave theory are
simple, locally causal, classical dymamical systems
defined by an update rule that specifies how the
unit updates its internal state based on an incoming message.
Such a unit exhibits an elementary form of learning capability
and operates deterministically and is therefore called a DLM~\cite{RAED05b}.

As dynamical systems, DLMs are interesting in their own right,
in particular because they may be used for applications that have no relation
to physics~\cite{RAED05b}.
In appendix A, we introduce them in a context-free manner.

\section{Simulation model}\label{sec5}

In this section, we give a detailed description of the EBCM of basic optical components,
the interface between to media and single-photon detectors.
For simplicity of presentation, we consider materials
for which the permittivity is real and the permeability is one,
that is we consider ideal dielectrics only.

\subsection{Messenger and message}

As we want to construct an EBCM that reproduces the results of \MT\
and the plane wave is a central concept in \MT~\cite{BORN64},
it makes sense to define the messenger and the message such that there is
a one-to-one mapping onto the properties of the plane wave.
Therefore, we briefly review the properties of plane electromagnetic waves
propagating in a homogeneous medium and then
show how to construct an EBCM that reproduces these properties
without using wave mechanics.

The EBCM that we introduce in this paper is not a complete
event-based model for all electrodynamic phenomena.
The focus of this paper is on optics experiments
in which the indivisabile nature of the photons
is of central importance.
In accordance with the classical theory of optics~\cite{BORN64},
for simplicity, we therefore ignore the
interaction of electromagnetic radiation with
the magnetic degrees of freedom of the materials.
It is however not difficult to incoporate these interactions
in the EBCM: All that is needed is to add more data to
the message carried by the messengers and to
define the appropriate transformation rules.

\subsubsection{Plane waves in Maxwell's theory}

A plane wave is fully specified by its wave vector $\mathbf{q}$
and the amplitudes and phases of the electric field components that are in the plane orthogonal to $\mathbf{q}$.
These amplitudes and phases determine the intensity and polarization of the electromagnetic wave.
The amplitudes oscillate in time with a common frequency $f$ which is related
to the magnitude of the wave vector $q=|\mathbf{q}|$ and the phase velocity $v$ by $q=2\pi f/v$, see Ref.~\cite{BORN64}.
In vacuum $v$ equals the speed of light $c$.
If the plane wave propagates through a dielectric material, its velocity is
given by $v=c/n$ where $n$ is a real number, the index of refraction of the material.
Note that $v$ and $n$ are determined indirectly via the law of refraction~\cite{BORN64}.

For a plane wave traveling along the $z$-direction, the amplitudes of the non-zero electric field components
can be written as~\cite{BORN64}
\begin{eqnarray}
E_1(\mathbf{r},t)&=&a_1 \cos (\tau+\delta_1)
\nonumber \\
E_2(\mathbf{r},t)&=&a_2 \cos (\tau+\delta_2)
,
\label{mess0}
\end{eqnarray}
where the $1$-$2$ coordinate system can be chosen freely as long as it lies in the plane orthogonal to $z$ and
$\tau=2\pi f t - \mathbf{q}\cdot \mathbf{r}$ where $\mathbf{r}$ denotes the position in 3D space.
The prefactors $a_1$ and $a_2$ determine the intensity of the plane wave
and together with the phase difference $\delta=\delta_1-\delta_2$
also determine its polarization~\cite{BORN64}.

In \MT, the amplitudes $a_1$ and $a_2$ determine the light intensity whereas
in an EBCM, each messenger is assumed to contributes one unit of light to the total light intensity.
Let us therefore rescale the fields $E_1$ and $E_2$ such that the average power of the $E$-field
is one half. We have
\begin{eqnarray}
\frac{E_1(\mathbf{r},t)}{\sqrt{a_1^2+a_2^2}}&=&\sin\xi \cos (\tau+\delta_1)
\nonumber \\
\frac{E_2(\mathbf{r},t)}{\sqrt{a_1^2+a_2^2}}&=&\cos\xi \cos (\tau+\delta_2)
,
\label{mess1}
\end{eqnarray}
where $\sin\xi=a_1/\sqrt{a_1^2+a_2^2}$ and $\cos\xi=a_2/\sqrt{a_1^2+a_2^2}$.
Expressing the Stokes parameters~\cite{BORN64} $s_0=a_1^2+a_2^2$,
$s_1=a_1^2-a_2^2$, $s_2=2a_1^2a_2^2\cos\delta$, $s_3=2a_1^2a_2^2\sin\delta$
in terms of the angles $\xi$ and $\delta$ yields
$s_1/s_0=-\cos2\xi$, $s_2/s_0=\sin2\xi\cos\delta$, $s_3/s_0=\sin2\xi\sin\delta$,
demonstrating that Eq.~(\ref{mess1}) is just another representation of a plane monochromatic electromagnetic wave.

\subsubsection{Definition of messenger and message}

The particle is regarded as a messenger, traveling with velocity $v$ in the direction $\mathbf{q}/q$.
Each messenger carries with it two harmonic oscillators that vibrate with frequency $f$.

It may be tempting, but it is definitely wrong, to regard the messenger+message as a plane wave
with wave vector $\mathbf{q}$, the oscillators being the two electric field
components in the plane orthogonal to $\mathbf{q}$.
As there is no communication/interaction between the messengers
there can be no wave equation (i.e. no partial differential equation)
that enforces a relation between the messages carried by different messengers.
Indeed, the oscillators carried by a messenger never interact with the oscillators
of another messenger, hence the motion of these pairs of oscillators is not governed by a wave equation.
Naively, one might imagine the oscillators tracing out a wavy pattern in space
as they travel though space.
However, as there is no relation between the times at which the messengers leave the source,
it is impossible to characterize all these traces by a field that
depends on one set of space-time coordinates, as required for a wave theory.

There are many different but equivalent ways to define the message.
As in \MT\ and \QT, it is convenient (though not essential) to work with complex-valued
vectors, that is with messages represented by two-dimensional unit vectors
\begin{equation}
\mathbf{y}=
\left(\begin{array}{c}
        e^{i\psi^{(1)}} \sin\xi\\
        e^{i\psi^{(2)}} \cos\xi
\end{array}\right).
\label{mess2}
\end{equation}
where $\psi^{(i)}=2\pi f t +\delta_i$ for $i=1,2$.
Note that, unlike in the case of waves, there is no $\mathbf{q}\cdot \mathbf{r}$ contribution to $\psi^{(i)}$.
The angle $\xi$ determines the relative magnitude of the two components, which we call the polarization
of the message $\mathbf{y}$.

A messenger with message $\mathbf{y}$ at time $t$ and position $\mathbf{r}$
that travels along the direction $\mathbf{q}$ during a time interval $t'-t$,
changes its message according to
\begin{equation}
\psi^{(i)}\leftarrow\psi^{(i)}+\phi\quad,\quad i=1,2
,
\label{mess3}
\end{equation}
where $\phi=2\pi f (t'-t)$.

From Eq.~(\ref{mess3}), as the messenger moves, it changes its
message by applying to the
vectors $(\cos \psi^{(1)},\sin \psi^{(1)})$ and $(\cos \psi^{(2)},\sin \psi^{(2)})$
the same plane rotation
\begin{equation}
R(\phi)=
\left(\begin{array}{cc}
\cos\phi &         {-}\sin\phi\\
\sin\phi & \phantom{-}\cos\phi
\end{array}\right)
.
\label{mess4}
\end{equation}
This suggests that we may view these two-component vectors as
the coordinates of two local oscillators, carried along by the messengers.
In this pictorial description, the messenger encodes its time-of-flight
in these two oscillators.

\subsubsection{Remarks}

Recall that in \MT, the energy of the electromagnetic field is encoded
in the amplitudes of the wave components.
In contrast, in the event-based corpuscular approach,
the energy of the electromagnetic field is encoded in the amount of particles that crosses
an unit area per unit of time.

The representation of the messages as a vector with two complex-valued components is
convenient, not only to describe the propagation in a homogenous
medium but, as we will see later, also to describe the effect
of for instance a beam splitter, a polarizing beam splitter and so on.
In essence, the interaction of the messenger with a material
will change the message $y$ by applying some transformation to it.
However, there is no rational argument to use the representation
Eq.~(\ref{mess2}) other than that it is convenient.
Any other representation, such as the ones used in our earlier papers work equally well.

When a messenger is created, its message needs to be initialized, that is we have
to specify the three angles $\psi^{(1)}$, $\psi^{(2)}$, and $\xi$.
This specification depends on the kind of light source we want to simulate.
For instance, to simulate a coherent light source, the three angles should be the same
for all messengers that are being created.
Other choices are discussed when we consider specific experiments.

\subsection{Interface between two dielectric media}

\begin{figure}[t]
\includegraphics[width=6cm]{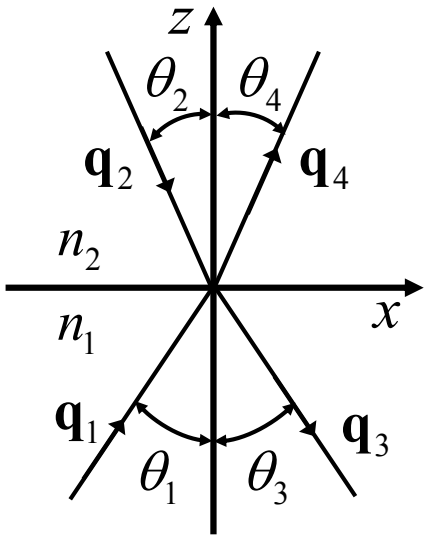}
\caption{%
Refraction and reflection at a boundary between two dielectric materials with indices
of refraction $n_1$ and $n_2$.
The vectors $\mathbf{q}_1$, $\mathbf{q}_2$, $\mathbf{q}_3$, and $\mathbf{q}_4$ lie in the $x$--$z$ plane.
}
\label{figinterface}
\end{figure}

The reflection and transmission of light by a boundary that separates
two homogeneous media of different optical properties is one of the basic phenomena in optics.
We briefly review the wave theoretical treatment of this problem
and explain how we construct the corresponding EBCM.

\subsubsection{Wave theory}

In \MT, assuming incident plane waves from both sides of the interface (see Fig.~\ref{figinterface}),
the directions and amplitudes of the reflected
and transmitted plane wave follow from conservation of energy and
the continuity of the tangential field components at the boundary,
yielding Snell's law and Fresnel's formulas, respectively.

In \MT, Snell's law follows from the kinematic properties of the waves.
Conservation of momentum in the $x$-direction implies $q^{(1)}_x = q^{(2)}_x=q^{(3)}_x=q^{(4)}_x$.
Using the relation between the frequency $f$ of the oscillations and the length of the wave vectors
we find
\begin{equation}
\frac{2\pi f}{c}=\frac{q^{(1)}}{n_1} = \frac{q^{(2)}}{n_2}= \frac{q^{(3)}}{n_1}= \frac{q^{(4)}}{n_2}
,
\label{int0}
\end{equation}
where
\begin{equation}
q^{(i)} = \sqrt{(q^{(i)}_x)^2 +(q^{(i)}_z)^2}\quad,\quad i=1,2,3,4
,
\label{int1}
\end{equation}
from which
\begin{equation}
n_1\frac{q^{(1)}_x}{q^{(1)}} = n_2\frac{q^{(2)}_x}{q^{(2)}}=n_1\frac{q^{(3)}_x}{q^{(3)}}=n_2\frac{q^{(4)}_x}{q^{(4)}}
,
\label{int2}
\end{equation}
which is Snell's law in its more familiar form
\begin{equation}
n_1\sin\theta_1 = n_2\sin\theta_2 =n_1\sin\theta_3=n_2\sin\theta_4
.
\label{int3}
\end{equation}
For later use, we introduce the symbols
$\widehat q^{(i)}=\mathbf{q}^{(i)}_z/{q}^{(i)}=(-1)^i\cos\theta_i$ for $i=1,2,3,4$,
$n_3=n_1$ and $n_4=n_2$.
Note that we also have $\theta_3=\theta_1$, $\theta_4=\theta_2$,
$\widehat q^{(3)}=-\widehat q^{(1)}$, and $\widehat q^{(2)}=-\widehat q^{(4)}$.
The kinematic properties of the particles in the EBCM
are assumed to be the same as those of the waves in \MT.

In \MT\ the dynamics properties are contained in the boundary conditions
of the electromagnetic fields at the interface.
If $A_i$ denotes the electric-field amplitude of the plane wave with
wave vector $\mathbf{q}^{(i)}$, these boundary conditions enforce the relations
\begin{equation}
\left(\begin{array}{c}
A_3\\
A_4
\end{array}\right)
=
\left(\begin{array}{lr}
r_{12} &t_{21}\\
t_{12}&-r_{12}
\end{array}\right)
\left(\begin{array}{c}
A_1\\
A_2
\end{array}\right)
.
\label{int4}
\end{equation}
The reflection and transmission coefficients are~\cite{BORN64}
\begin{eqnarray}
r_{12}&=&\frac{n_1\widehat q^{(1)}-n_2\widehat q^{(4)}}{n_1\widehat q^{(1)}+n_2\widehat q^{(4)}}
\nonumber \\
t_{12}&=&\frac{2n_1\widehat q^{(1)}}{n_1\widehat q^{(1)}+n_2\widehat q^{(4)}}
\nonumber \\
t_{21}&=&\frac{2n_2\widehat q^{(4)}}{n_1\widehat q^{(1)}+n_2\widehat q^{(4)}}
,
\label{int5}
\end{eqnarray}
for $S$-polarized waves and
\begin{eqnarray}
r_{12}&=&\frac{n_1\widehat q^{(4)}-n_2\widehat q^{(1)}}{n_1\widehat q^{(4)}+n_2\widehat q^{(1)}}
\nonumber \\
t_{12}&=&\frac{2n_1\widehat q^{(1)}}{n_1\widehat q^{(4)}+n_2\widehat q^{(1)}}
\nonumber \\
t_{21}&=&\frac{2n_2\widehat q^{(4)}}{n_1\widehat q^{(4)}+n_2\widehat q^{(1)}}
,
\label{int6}
\end{eqnarray}
for $P$-polarized waves.

In \MT\ the wave amplitudes are only convenient calculational tools:
Observable effects are related to the energy current per unit area~\cite{BORN64}, the
Poynting vectors $\mathbf{S}^{(i)}=cn_i\mathbf{q}^{(i)}|A_i|^2/4\pi |\mathbf{q}^{(i)}|$.

It is not difficult to rewrite Eq.~(\ref{int4}) such that it directly relates to physical observables.
Define the ``energy-current amplitude'' $\widetilde A_i=(cn_i|\widehat{q}^{(i)}|/4\pi)^{1/2} A_i$
and the transformation matrix $R$ by
\begin{equation}
R=
\left(\begin{array}{lr}
(cn_1\widehat{q}^{(1)}/4\pi)^{-1/2} &0\\
0&(cn_2\widehat{q}^{(2)}/4\pi)^{-1/2}
\end{array}\right)
.
\label{int7}
\end{equation}
Then, from Eq.~(\ref{int4}) it follows that
\begin{eqnarray}
\left(\begin{array}{c}
\widetilde A_3\\
\widetilde A_4
\end{array}\right)
&=&
R^{-1}
\left(\begin{array}{lr}
r_{12} &t_{21}\\
t_{12}&-r_{12}
\end{array}\right)
R
\left(\begin{array}{c}
\widetilde A_1\\
\widetilde A_2
\end{array}\right)
,
\label{int8}
\nonumber \\
&=&
\left(\begin{array}{lr}
\widetilde r & \widetilde t\\
\widetilde t&-\widetilde r
\end{array}\right)
\left(\begin{array}{c}
\widetilde A_1\\
\widetilde A_2
\end{array}\right)
\equiv
\widetilde T
\left(\begin{array}{c}
\widetilde A_1\\
\widetilde A_2
\end{array}\right)
,
\label{int9}
\end{eqnarray}
where $\widetilde r$ and $\widetilde t$ are given by
\begin{eqnarray}
\widetilde r_S&=&\frac{n_1\widehat q^{(1)}-n_2\widehat q^{(4)}}{n_1\widehat q^{(1)}+n_2\widehat q^{(4)}}
\nonumber \\
\widetilde t_S&=&\frac{2\sqrt{n_1n_2\widehat q^{(1)}\widehat q^{(4)}}}{n_1\widehat q^{(1)}+n_2\widehat q^{(4)}},
,
\label{int10}
\end{eqnarray}
and
\begin{eqnarray}
\widetilde r_P&=&\frac{n_1\widehat q^{(4)}-n_2\widehat q^{(1)}}{n_1\widehat q^{(4)}+n_2\widehat q^{(1)}},
\nonumber \\
\widetilde t_P&=&\frac{2\sqrt{n_1n_2\widehat q^{(1)}\widehat q^{(4)}}}{n_1\widehat q^{(4)}+n_2\widehat q^{(1)}}
,
\label{int11}
\end{eqnarray}
in the case of $S$-polarized and $P$-polarized waves, respectively.
From Eqs.~(\ref{int6}), (\ref{int10}) and (\ref{int11}) it follows directly that $\widetilde r^2+\widetilde t^2=1$
and that the matrix $\widetilde T$ is unitary (with determinant minus one).
As $\widetilde T$ is unitary we have $|\widetilde A_3|^2+ |\widetilde A_4|^2=|\widetilde A_1|^2+ |\widetilde A_2|^2$,
that is the total energy of the incident waves is equal to the total energy of the outgoing waves, as it should be.

Clearly, in an EBCM of refraction and reflection at a dielectric, lossless interface,
there can be no loss of particles: An incident particle must either bounce back from or pass through the interface.
If the EBCM is to reproduce the results of \MT\,
the boundary conditions on the wave amplitudes in \MT\ must translate
into a rule that determines how a particle bounces back or crosses the interface.
But in the EBCM there are only particles, no wave amplitudes.
However, as we have seen in \MT\ the wave energies, not the wave amplitudes, appear in conservation laws.
This suggests that in an EBMC, we should use the matrix $\widetilde T$ to transform
the message carried by the messenger.

\subsubsection{EBCM of the interface}\label{ebcmint}

\begin{figure}[t]
\begin{center}
\includegraphics[width=8cm]{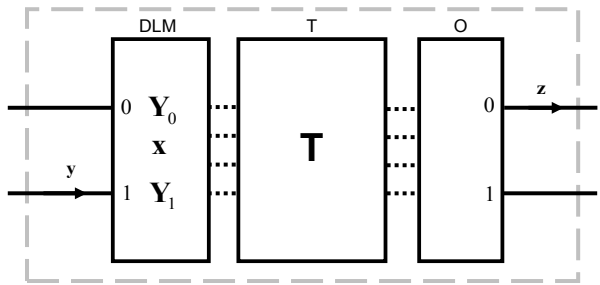}
\caption{Diagram of a DLM-based processing unit that performs an event-based simulation of
optical components.
The processing unit consists of three stages: An input stage (DLM), a transformation stage (T) and an output stage (O).
The solid lines represent the input and output ports of the device.
The presence of a message is indicated by an arrow on the corresponding port line.
The dashed lines indicate the data flow within the unit.
The transformation matrix ${\bf T}$ is component specific.
}
\label{figmachine0}
\end{center}
\end{figure}

We now have all ingredients to construct the processing unit that performs
the event-by-event, corpuscular simulation of refraction and reflection of light
at a dielectric, lossless interface.

The processing unit has the generic structure, depicted in Fig.~\ref{figmachine0},
consisting of an input stage (DLM),
a transformation stage (T), and an output stage (O)~\cite{RAED05d,RAED05b,RAED05c,MICH05}.
There are two input and two output ports labeled by $k=0,1$.
Refering to Fig.~\ref{figinterface}, input port $k=0,1$ accepts messengers travelling along
directions $\mathbf{q}^{(1)}/{q}^{(1)}$ and $\mathbf{q}^{(2)}/{q}^{(2)}$, respectively
(recall that at any time, only one messenger arrives at either port 0 or 1).
Likewise, a messenger leaves in the directions $\mathbf{q}^{(3)}/{q}^{(3)}$ and $\mathbf{q}^{(4)}/{q}^{(4)}$
through output port $k=0,1$, respectively.

For simplicity, we assume that the incoming messenger is constructed such that
its first oscillator vibrates in the plane that is orthogonal to the plane of incidence (the $xz$-plane in Fig.~\ref{figinterface})
while its second oscillator vibrates in the plane that is parallel to the plane of incidence,
corresponding to $S$ and $P$ polarized plane waves, respectively.
Note that this assumption merely amounts to a convenient choice of the coordinate system.

\subsubsection{Input stage}
The DLM receives a message on either input port $0$ or $1$, never on both ports simultaneously.
If we represent the arrival of a messenger at port 0 (1) by
the vectors ${\bf v}=(1,0)$ or ${\bf v}=(0,1)$, it is clear that
we can employ the DLM of Section~\ref{DLMC}, second case, to estimate the relative frequencies
with which the messengers arrive on ports 0 or 1, respectively.

According to Section~\ref{DLMC}, a DLM that is capable of performing this task
should have an internal vector $\mathbf{x}=( x_{0},x_{1})$,
where $x_{0}+x_{1}=1$ and $x_{k}\geq 0$ for all $k=0,1$.
In addition to the internal vector $\mathbf{x}$,
the DLM needs to have two sets of two registers
$\mathbf{Y}_{k}=(Y_{k,1},Y_{k,2})$
to store the last message $\mathbf{y}$ that arrived at port $k$.
Thus, the DLM has storage for exactly ten real-valued numbers.

Upon receiving a messenger at input port $k$, the DLM performs the following steps:
It copies the elements of message $\mathbf{y}$ in its internal register $\mathbf{Y}_k$
\begin{equation}
\mathbf{Y}_k\leftarrow\mathbf{y}
\label{int21}
\end{equation}
while leaving $\mathbf{Y}_{1-k}$ unchanged
and replaces its internal vector according to
\begin{equation}
\mathbf{x}\leftarrow\gamma \mathbf{x}+( 1-\gamma ) \mathbf{v},
\label{int20}
\end{equation}
where $0\le\gamma <1$.
Note that each time a messenger arrives at one of the input ports,
the DLM replaces the values of the internal vector $\mathbf{x}$
and of the ones in the registers $\mathbf{Y}_{k}$
by overwriting the old ones.
It does not store all the messages, but only two!

\subsubsection{Transformation stage}

The second stage (T) accepts a message from the input stage, and transforms it into a new message.
From the description of the input stage, it is clear that
the internal registers $\mathbf{Y}_0$ and $\mathbf{Y}_1$
contain the last message that arrived on input port 0 and 1 respectively.
First, this data is combined with the data of the internal vector
$\mathbf{x}$, the components of which converge (after many events
have been processed) to the relative frequencies with which
the messengers arrive on port 0 and 1, respectively.
The output message generated by the input stage is
\begin{equation}
\left(\begin{array}{c}
        {Y}_{0,1}'\\
        {Y}_{1,1}'\\
        {Y}_{0,2}'\\
        {Y}_{1,2}'
\end{array}\right)
=
\left(\begin{array}{cccc}
        x_0^{1/2}&0&0&0\\
        0&x_1^{1/2}&0&0\\
        0&0&x_0^{1/2}&0\\
        0&0&0&x_1^{1/2}
\end{array}\right)
\left(\begin{array}{c}
        {Y}_{0,1}\\
        {Y}_{1,1}\\
        {Y}_{0,2}\\
        {Y}_{1,2}
\end{array}\right)
,
\label{trans0}
\end{equation}
Note that as $x_0+x_1=1$ and $\Vert\mathbf{Y}_{0}\Vert=\Vert\mathbf{Y}_{1}\Vert=1$,
we have $|{Y}_{0,1}'|^2 +|{Y}_{0,2}'|^2 +|{Y}_{1,1}'|^2 +|{Y}_{1,2}'|^2=1$.

Recall that in the EBCM, the number of incoming messengers with message $\mathbf{Y}_{k}$
represents the energy-density current of the corresponding plane wave, which by construction is
proportional to $|\mathbf{Y}_{k}'|^2$.
Therefore, according to Eq.~(\ref{int9}), the outgoing energy-density currents are given by
\begin{equation}
\left(\begin{array}{cc}
        {Z}_{0,1}\\
        {Z}_{1,1}\\
        {Z}_{0,2}\\
        {Z}_{1,2}
\end{array}\right)
=
\mathbf{T}
\left(\begin{array}{c}
        {Y}_{0,1}'\\
        {Y}_{1,1}'\\
        {Y}_{0,2}'\\
        {Y}_{1,2}'
\end{array}\right)
,
\label{trans1}
\end{equation}
where the transformation matrix $\mathbf{T}$ is given by
\begin{equation}
\mathbf{T}=
\left(\begin{array}{cccc}
        \widetilde r_S&\phantom{-}\widetilde t_S&0&0\\
        \widetilde t_S&-\widetilde t_S&0&0\\
        0&0&\widetilde r_P&\phantom{-}\widetilde t_P\\
        0&0&\widetilde t_P&-\widetilde r_P
\end{array}\right)
,
\label{trans2}
\end{equation}
and the matrix elements are given by Eqs.~(\ref{int10}) and (\ref{int11}).

\subsubsection{Output stage}

The output stage (O) uses the data provided by the transformation stage (T)
to decide on which of the two ports it will send out a messenger (representing a photon).
The rule is very simple:
We compute $z=|{Z}_{1,1}|^2+|{Z}_{1,2}|^2$ and select the output port $\widehat k$ by the rule
\begin{equation}
\widehat k = \Theta(z - r)
,
\label{out0}
\end{equation}
where $\Theta(.)$ is the unit step function
and the $0\leq r <1$ is a uniform pseudo-random number (which is different for each messenger processed).
The messenger leaves through port $\widehat k$ carrying the message
\begin{equation}
\mathbf{z} =
\frac{1}{\sqrt{|{Z}_{\widehat k,1}|^2+|{Z}_{\widehat k,2}|^2}}
\left(\begin{array}{c}
        {Z}_{\widehat k,1}\\
        {Z}_{\widehat k,2}
\end{array}\right)
,
\label{out1}
\end{equation}
which, for reasons of internal consistency, is a unit vector.

\subsubsection{Remarks}

The internal vector of the DLM can be given physical meaning: It represents
the polarization vector of the charge distribution of an atom.
The update rule Eq.~(\ref{int20}) defines the equation of motion
of this vector.
In essence, the EBCM is a simplified version of the
classical Newtonion Lorentz model for the response of the polarization
of an atom/molecule to the applied electric field~\cite{BORN64},
with one important difference: It describes the interaction
of a single photon with the atom.
The update rule Eq.~(\ref{int20}) is not the only rule
which yields an EBCM that reproduces the results of \MT\ (see Section~\ref{validation}).
The question of the correctness of the update rule can only be settled by
a new type of experiment that addresses this specific question.

The use of pseudo-random numbers to select the output port is not essential~\cite{RAED05b}:
We generally use pseudo-random numbers to mimic the apparent unpredictability
of the experimental data only.
From a simulational point of view, there is nothing special about using pseudo-random numbers.
On a digital computer, pseudo-random numbers are generated by deterministic processes anyway
hence instead of a uniform pseudo-random number generator, any algorithm
that selects the output port in a systematic manner might be employed as well~\cite{RAED05b},
as long as the zero'a and one's occur with a ratio determined by $z$.
For instance, we could use the DLM described in Section~\ref{DLMB}.
This will change the order in which messengers are being processed but
the stationary state averages are the same.

\subsection{Single-photon detectors}\label{SPD}

Ideally, each photon that hits a single-photon detector should simply produce a signal, a click.
This idealized model is used in virtually all \QT\  modeling for very good reasons:
Quantum theory postulates that the observed intensity, the average over many events,
recorded by a detector is proportional to the square
of the absolute value of the complex-valued wave amplitude.
Likewise, Maxwell's theory interprets the intensity of light
as the energy flux (Poynting vector) per unit of time through
a unit area orthogonal to the directions of the electric and magnetic
field vector~\cite{BORN64}.
Part of the predictive power of both wave theories
stems from these postulates as they allow the calculation of intensities
completely independent of the details of how the intensity is actually measured.
In jargon, these theories are noncontextual~\cite{HOME97}.

It is generally accepted that a description on the level of individual events necessarily
entails contextuality~\cite{HOME97}.
Therefore, in plain words, a description that aims to go beyond
 \QT\ or \MT, that is a cause-and-effect description on the level of individual events,
requires a detailed specification of the process of measurement itself~\cite{HOME97}.
Obviously, such a specification should capture the essence of a real detection process.

In reality, photon detection is the result of a complicated
interplay of different physical processes~\cite{HADF09}.
Therefore we briefly review some of the basic features of these processes
before we discuss the detector model that we employ in our simulations.

\subsubsection{Generalities}
In essence, a light detector consists of material that absorbs light.
The electric charges that result from the absorption process are then amplified,
chemically in the case of a photographic plate or electronically in the case of photo diodes or photomultipliers.
In the case of photomultipliers or photodiodes,
once a photon has been absorbed (and its energy ``dissipated'' in the detector material)
an amplification mechanism (which requires external power/energy) generates
an electric current (provided by an external current source)~\cite{GARR09,HADF09}.
The resulting signal is compared with a threshold that is set
by the experimenter and the photon is said to have been detected
if the signal exceeds this threshold~\cite{GARR09,HADF09}.
In the case of photographic plates, the chemical process that occurs when photons
are absorbed and the subsequent chemical reactions that
renders visible the image serve similar purposes.

In the wave-mechanical picture, the interaction between the incident electric field ${\mathbf E}$ and
the material takes the form ${\mathbf P}\cdot{\mathbf E}$, where ${\mathbf P}$ is the polarization vector of the material~\cite{BORN64}.
Treating this interaction in first-order perturbation theory, the detection probability reads
$P_\mathrm{detection}(t)=\int^{t}_{0}\int^{t}_{0}\langle\langle{\mathbf E^{T}(t')}\cdot{\mathbf K(t'-t'')}\cdot{\mathbf E(t'')}
\rangle\rangle dt'dt''$ where ${\mathbf K(t'-t'')}$ is a memory kernel depending on the material only and
$\langle\langle.\rangle\rangle$ denotes the average with respect to the initial state of the electric field~\cite{BALL03,GARR09}.
Both the constitutive equation~\cite{BORN64} ${\mathbf P}(\omega)=\chi(\omega){\mathbf E}(\omega)$ as well as the expression
for $P_\mathrm{detection}(t)$ show that the detection process involves some kind of memory,
as is most evident in the case of photographic films before development.
It is important to take note that for the integration over $t'$ and $t''$ to yield physically meaningful results,
within the context of a wave theory,
the interaction interval $[0,t]$ should extend over many periods of the wave~\cite{BORN64}.

An event-based model for the detector cannot be ``derived'' from  \QT\
simply because  \QT\ has nothing to say about individual events
but predicts the frequencies of their observation only~\cite{HOME97}.
Therefore, any model for the detector that operates on the level of single events
must necessarily appear as ``ad hoc'' from the viewpoint of  \QT.
The event-based detector models that we describe in this paper should not be regarded
as realistic models for say, a photomultiplier or a photographic plate and the
chemical process that renders the image.
In particular, there is no need to account for the presence of a threshold (electrical or chemical)
that is essential for real photon-detection systems to function properly~\cite{GARR09,HADF09}.
In the spirit of Occam's razor, these very simple event-based models capture the salient features
of ideal (i.e. 100\% efficient)
single-photon detectors without making reference to the solution of a wave equation or \QT.

\begin{figure}[t]
\begin{center}
\includegraphics[width=8cm]{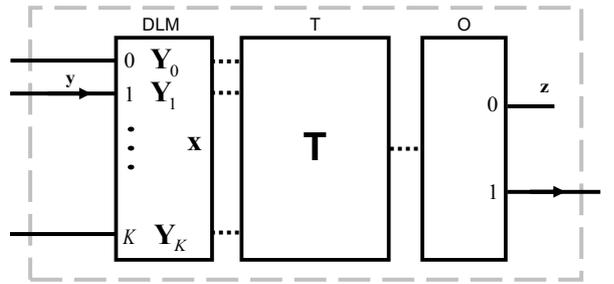}
\caption{%
Diagram of the event-based detector model defined by Eqs.~(\ref{det0}), (\ref{det1}), and (\ref{det2}).
The dashed line indicates the data flow within the processing unit.
}%
\label{figdetector}
\end{center}
\end{figure}

\subsubsection{EBCM of a detector}\label{detector}

Photon detectors, such as a photographic plate of CCD arrays, consist of many identical detection units
each having a predefined spatial window in which they can detect photons.
In what follows, each of these identical detection units will be referred to as a detector.
By construction, these detector units operate completely independently from and also do not communicate with each other.

Here we construct a processing unit that acts as a detector for individual messages.
Instead of differentiating between only two different directions as in the case of the EBCM of an interface for instance,
a detector should be able to process messengers that come from many different directions.
The schematic diagram depicted in Fig.~\ref{figdetector} shows that this processing unit has the same structure
as the general processing unit, see Fig.~\ref{figmachine0} except that it may have more than two inputs ports.
The number of input ports is denoted by $N_p=K+1$.

\subsubsection{Input stage}
Representing the arrival of a messenger at port $0\le k\le K$ by
the vector ${\bf v}=(v_0,\ldots,v_K)^T$ with $v_i=\delta_{i,k}$
the internal vector is updated according to the rule
\begin{equation}
\mathbf{x} \leftarrow \widehat\gamma \mathbf{x} + (1-\widehat\gamma) \mathbf{v}
,
\label{det0}
\end{equation}
where $\mathbf{x}=(x_0,\ldots,x_{K})^T$, $\sum_{k=1}^K x_k=1$, and $0\le\widehat\gamma<1$.
The elements of the incoming message $\mathbf{y}$ are written in internal register $\mathbf{Y}_k$
\begin{equation}
\mathbf{Y}_k\leftarrow\mathbf{y}
,
\label{det1}
\end{equation}
while all the other $\mathbf{Y}_{i}$ ($i\not=k$) registers remain unchanged.
Thus, each time a messenger arrives at one of the input ports, say $k$,
the DLM updates all the elements of the internal vector $\mathbf{x}$,
overwrites the data in the register $\mathbf{Y}_{k}$
while the content of all other $\mathbf{Y}$ registers remains the same.

\subsubsection{Transformation stage}

The output message generated by the transformation stage is
\begin{equation}
\mathbf{T}=\mathbf{x}\cdot\mathbf{Y}=\sum_{k=0}^K x_k \mathbf{Y}_k
,
\label{det2}
\end{equation}
which is a complex-valued two-component vector, similar to a message $\mathbf{y}$.

\subsubsection{Output stage}

As in all previous event-based models for the optical components,
the output stage (O) generates a binary output signal $\widehat k=0,1$
but unlike components such as a parallel plate, the output message does not represent a photon:
It represents a ``no click'' or ``click'' if $\widehat k=0$ or $\widehat k=1$, respectively.
To implement this functionality, we define (compare with Eq.~(\ref{out0}) which is essentially the same)
\begin{equation}
\widehat k = \Theta(|\mathbf{T}|^2- r)
,
\label{det3}
\end{equation}
where $\Theta(.)$ is the unit step function and $0\leq r <1$
are uniform pseudo-random numbers (which are different for each event).
The parameter $0\le \widehat \gamma<1$ can be used to control the operational mode of the unit.
From Eq.~(\ref{det3}) it follows that the frequency of $\widehat k=1$ events depends
on the length of the internal vector $\mathbf{T}$.

Note that in contrast to experiment, in a simulation, we could register both the $\widehat k=0$ and $\widehat k=1$ events.
Then the sum of the $\widehat k=0$ and $\widehat k=1$ events is equal to the number of input messages.
In real experiments, only $\widehat k=1$ events are taken as evidence that a photon has been detected.
Therefore, we define the total detector count by
\begin{equation}
N_{\mathrm{count}}=\sum^{K}_{l=1}\widehat k_l,
\label{det4}
\end{equation}
where $K$ is the number of messages received and $l$ labels the events.
In words, $N_{\mathrm{count}}$ is the total number of one's generated by the detector unit.

\subsubsection{Detection efficiency}

In general, the detection efficiency is defined as the overall probability of registering a count if a
photon arrives at the detector~\cite{HADF09}.
One method to measure the detection efficiency is to use a single-photon point source, placed far away
from a single detector~\cite{HADF09}.
In an EBCM of such an experiment, all the messengers that reach the detector will approximately
have the same direction, implying that these messengers arrive at this detector at the same input port, say $k$.
As explained in Section~\ref{DLMC}, under these circumstances, $x_k$
converges exponentially to one.
Hence, after receiving a few photons, the detector clicks every time a photon arrives.
Thus, the detection efficiency, as defined for real detectors,
of the event-based detector model is very close to 100\%.

Comparing the number of ad hoc assumptions and unknown functions that enter typical  \QT\ treatments
of photon detectors~\cite{GARR09} with the two parameters $\widehat \gamma$ and $N_p$ of the event-based detector model,
the latter has the virtue of being extremely simple while providing a description of
the detection process at the level of detail, the single events, which is outside the scope of  \QT.

\section{Model validation}\label{validation}

We validate the EBCM by simulating some very basic optical experiments
such as reflection and refraction by a boundary between two dielectric materials
and by a plane parallel plate and the interference of two light beams.

\subsection{Reflection and refraction from an interface}\label{RRI}

\begin{figure}[t]
\begin{center}
\includegraphics[width=8cm]{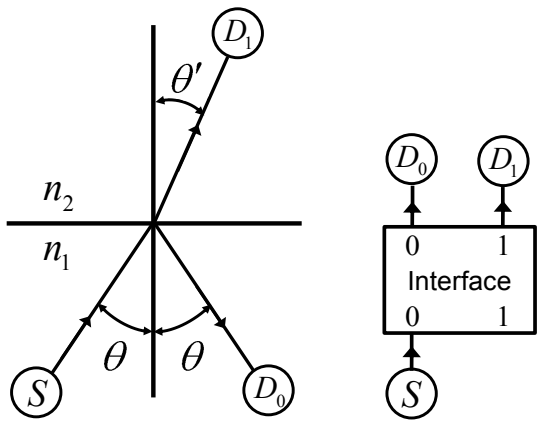}
\caption{%
Diagram of the experiment to measure the reflection and refraction from an interface (left)
and the diagram of the corresponding EBCM model (right).
In \MT\ the source emits plane waves and the amount of energy in the reflected and transmitted light
wave is recorded by detectors $D_0$ and $D_1$, respectively.
In the EBCM, the source emits particles which are either reflected or refracted by the interface
and the detectors $D_0$ and $D_1$ count the number of particles that are reflected or transmitted,
respectively.
}%
\label{expinterface}
\end{center}
\end{figure}

In Fig.~\ref{expinterface}, we show the diagram and its EBCM equivalent
of an experiment to measure the reflection and refraction by a boundary between two dielectric materials.
In the EBCM, the source sends messengers with their initial time-of-flight set to zero
to port 0 of the DLM-based processor, described in Section~\ref{ebcmint}.
The message consists of the vector $\mathbf{y}$ and the direction $\theta$
(or equivalently $\mathbf{q}_1$, see Fig.~\ref{figinterface}).
The DLM-based processor accepts the message, updates its internal state according
to the rules, given in Section~\ref{ebcmint}.
The result of this processing is that a messenger leaves the processor
via port 0 or (exclusive) 1, corresponding to a reflected or transmitted messenger, respectively.
The messenger then proceeds to either detector $D_0$ or $D_1$.
Each detector is also a DLM-based processor (recall that we impose the condition that
the same algorithms should be used in all experiments), as described in Section~\ref{detector}.
Counting the events produced by these detectors yields the reflectivity and transmissivity
of the interface.

\begin{figure}[t]
\begin{center}

\includegraphics[width=8cm]{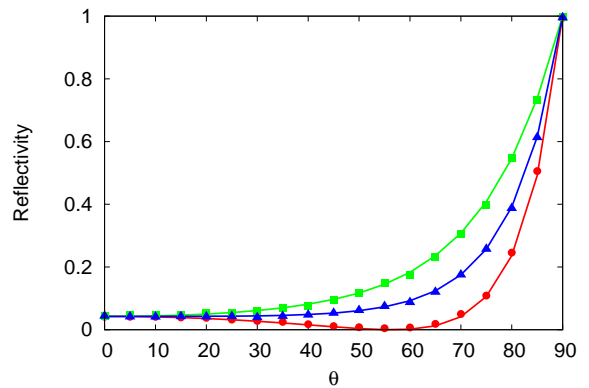}
\caption{%
EBCM simulation results (markers) for the reflectivity of an
interface between vacuum ($n_1=1$) and glass ($n_2=1.52$)
as a function of the angle of incidence of the incoming particles
(see also Fig.1.12 of Ref.~\cite{BORN64}).
The solid lines are the exact results of \MT~\cite{BORN64}.
Green markers and lines: $\xi=0$ ($S$-polarization);
Blue markers and lines: $\xi=\pi/4$ ($S$+$P$ polarization);
Red markers and lines:  $\xi=\pi/2$ ($P$-polarization).
Simulation parameters: $10^4$ events per marker, $\gamma=\widehat\gamma=0.99$ and $N_p=1$.
}%
\label{siminterface}
\end{center}
\end{figure}

In Fig.~\ref{siminterface} we compare EBCM simulation results with the predictions
of \MT~\cite{BORN64}. It is clear that there is excellent agreement, even though the number of emitted
events is small compared to the number of photons used in typical optics experiments.
Obviously, the EBCM passes this first test.

%

\subsection{Multiple-beam fringes with a plane-parallel plate}\label{MBI}

Light impinging on a transparant plane-parallel plate is multiply reflected at the plate boundaries,
a phenomenon that has many important applications.
The resulting interference effects have, according to current teaching of physics, no corpuscular analog.
Therefore, this system provides a stringent testcase for the EBCM approach.

\begin{figure}[t]
\begin{center}
\includegraphics[width=8cm]{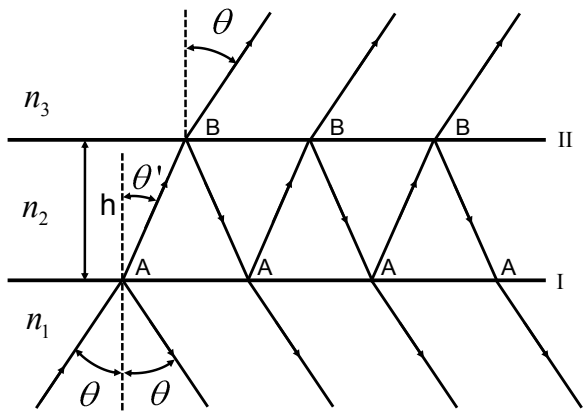}
\caption{%
Wave mechanical picture of multiple reflection in a plane-parallel plate~\cite{BORN64}.
}%
\label{expplate}
\end{center}
\end{figure}

In Fig.~\ref{expplate}, we present a schematic picture of the multiple interference that occurs
in the plane-parallel plate when the latter is illuminated by a plane wave.
We assume that the plate of thickness $h$ is a dielectric with an index of refraction $n_2$,
surrounded by dielectrica with indices of refraction $n_1$ and $n_3$.

\begin{figure}[t]
\begin{center}
\includegraphics[width=8cm]{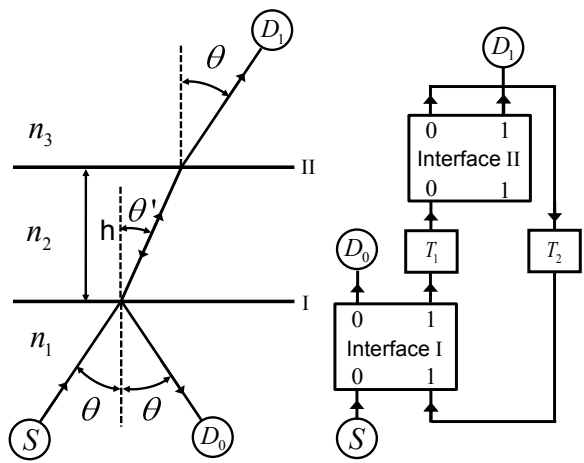}
\caption{%
Left: Diagram of the experiment to measure the reflection and refraction from a plane-parallel plate.
In \MT\ the source emits plane waves and the amount of energy in the reflected and transmitted light
wave is recorded by detectors $D_0$ and $D_1$, respectively.
Right: In the EBCM, the source emits particles which are either reflected or transmitted by the interface I.
In the latter case the particle travels to interface II where it is transmitted or reflected.
In the latter case the particle travels back to interface I.
The time-of-flights to travel from interface I to interface II and back again are $T_1$ and $T_2$, respectively.
The detectors $D_0$ and $D_1$ count the number of particles that are reflected or transmitted by the parallel
plate, respectively.
}%
\label{dlmplate}
\end{center}
\end{figure}

Refering to Fig.~\ref{expplate} and assuming an incident plane wave (as is done in \MT),
by translational invariance, all points labeled $A$ are equivalent.
Likewise, all points labeled $B$ are equivalent.
Therefore, we may replace the diagram of Fig.~\ref{expplate}
by the simpler one, shown in Fig.~\ref{dlmplate} (left).
Within the realm of \MT, this amounts to solving the wave equation for the plate directly,
without explicitly summing over all the contributions of the multiply reflected waves,
both approaches giving identical results~\cite{BORN64}.

The EBCM of this problem, simplified by exploiting the translation invariance,
is shown in Fig.~\ref{dlmplate} (right).
As in the case of the single interface, the source sends messengers with their initial time-of-flight set to zero
to port 0 of the DLM-based processor I that simulates interface I.
If processor I decides to send the messenger through port 0, the messenger
proceeds to detector $D_0$ where it generates a click and is counted as a reflected particle.
If the messenger leaves the processor I through port 1, the messenger travels to interface II
in straight line, increasing its time-of-flight by
$T_1=v_2^{-1}hf\cos\theta'$ where $n_2\sin\theta'=n_1\sin\theta$ (Snell's law) and $v_2=c/n_2$.
The algorithms executed by processors I and II are, of course, identical.
If processor II decides to send the messenger through port 1, the messenger
travels to detector $D_1$ where it generates a click and is counted as a transmitted particle.
Otherwise, the messenger returns to the first interface, increasing its time-of-flight by another amount of $T_2=v_2^{-1}hf\cos\theta'$,
and arrives at port 1 of processor I.
If processor I decides to send the messenger through port 0, the messenger
proceeds to detector $D_0$ where it generates a click and is counted as a reflected particle.
Otherwise the messenger is directed to processor II.
In this case, as is obvious from Fig.~\ref{dlmplate}(right),
the messenger makes one or more loops, traveling from processor I to II and back again.
This process mimics the multiple interference.
Only after a messenger has been detected by either $D_0$ or $D_1$,
the source may send a new messenger to port 0 of processor I.

\begin{figure}[t]
\begin{center}
\includegraphics[width=8cm]{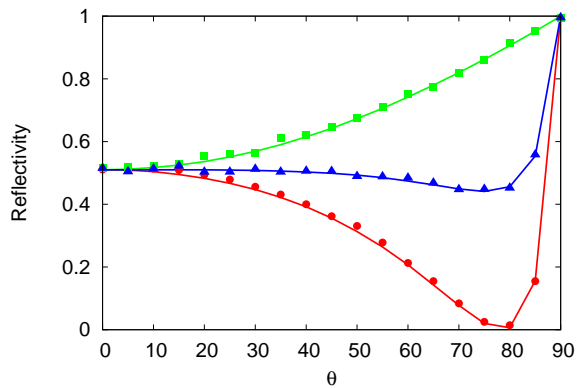}
\caption{%
EBCM simulation results (markers) for the reflectivity of
a parallel plate ($n_1=1$, $n_2=3$, $h=c/4fn_2$, $n_3=1.5$,
corresponding to a quarter-wave plate)
as a function of the angle of incidence of the incoming particles.
The solid lines are the exact results of \MT~\cite{BORN64}.
Green markers and lines: $\xi=0$ ($S$-polarization);
Blue markers and lines: $\xi=\pi/4$ ($S$+$P$ polarization);
Red markers and lines:  $\xi=\pi/2$ ($P$-polarization).
Simulation parameters: $10^4$ events per marker, $\gamma=\widehat\gamma=0.99$ and $N_p=1$.
}%
\label{simplate1}
\end{center}
\end{figure}

\begin{figure}[t]
\begin{center}
\includegraphics[width=8cm]{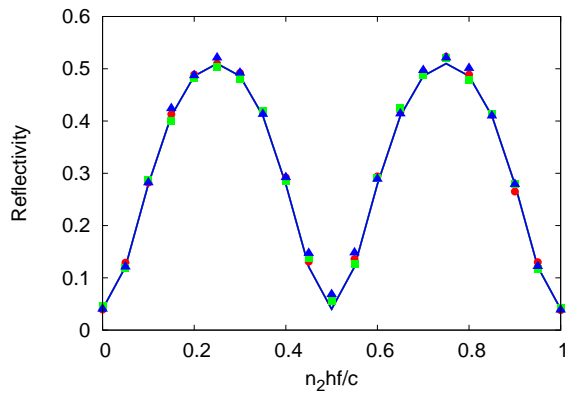}
\caption{%
EBCM simulation results (markers) for the reflectivity of
a parallel plate ($n_1=1$, $n_3=1.5$)
as a function of the optical thickness $n_2 h f/c$
for normal incidence of the incoming particles
(see also Fig.1.18 of Ref.~\cite{BORN64}).
The solid lines are the exact results of \MT~\cite{BORN64}.
For the legend and simulation parameters, see Fig.~\ref{simplate1}.
For normal incidence, markers and solid lines for different polarizations coincide.
}%
\label{simplate2}
\end{center}
\end{figure}

In Figs.~\ref{simplate1} and Figs.~\ref{simplate2}, we present EBCM simulation results and compare them with the predictions
of \MT~\cite{BORN64}.
It is clear that in all cases, there is excellent agreement, demonstrating that the EBCM of the interface can be used as a basic
building block for constructing optical components such as wave plates, beam splitters and the like.

%

\subsection{Two-beam interference experiments}\label{TBI}

In 1924, de Broglie introduced the idea that also matter, not just light, can exhibit wave-like properties~\cite{BROG25}.
This idea has been confirmed in various double-slit experiments with massive objects such as
electrons~\cite{JONS61,MERL76,TONO89,NOEL95}, neutrons~\cite{ZEIL88,RAUC00}, atoms~\cite{CARN91,KEIT91}
and molecules such as $C_{60}$ and $C_{70}$~\cite{ARND99,BREZ02}, all showing interference.
In some of the double-slit experiments~\cite{MERL76,TONO89,JACQ05}
the interference pattern is built up by recording individual clicks of the detectors.
The intellectual challenge is to explain how the detection of individual objects that do not interact with each other
can give rise to the interference patterns that are being observed.
According to Feynman, the observation that the interference patterns are built up event-by-event
is ``impossible, absolutely impossible to explain in any classical way
and has in it the heart of quantum mechanics''~\cite{FEYN65}.
In this section we demonstrate that this conclusion needs to be revised
by constructing an EBCM for the interference experiment with two light beams
that reproduces the results of \MT.

\subsubsection{Wave theory}
\begin{figure}[t]
\begin{center}
\includegraphics[width=8cm]{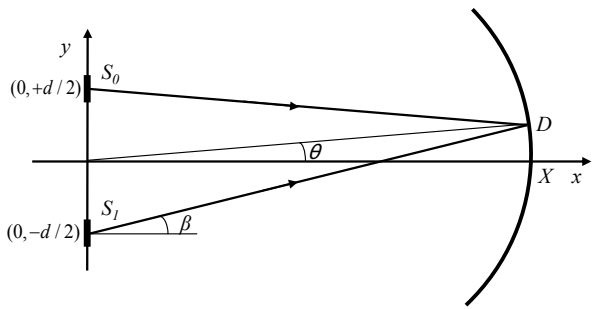}
\caption{%
Schematic diagram of the double-slit experiment with two light sources $S_0$ and $S_1$ of width $a$,
separated by a center-to-center distance $d$,
emitting light according to a uniform current distribution (see Eq.~(\ref{tbi1}))
and with a uniform angular distribution, $\beta$ denoting the angle.
The light is recorded by detectors $D$ positioned on a semi-circle with radius $X$.
The angular position of a detector is denoted by $\theta$.
}
\label{exptbi}
\end{center}
\end{figure}

We consider the simple experiment sketched in Fig.~\ref{exptbi}.
The sources $S_0$ and $S_1$ are lines of length $a$, separated by a center-to-center distance $d$.
These sources emit coherent, monochromatic light according to a uniform current distribution, that is
\begin{equation}
J(x,y)=\delta(x)\left[\Theta(a/2-|y-d/2|)+\Theta(a/2-|y+d/2|)\right]
,
\label{tbi1}
\end{equation}
where $\Theta(.)$ denotes the unit step function.
In the Fraunhofer regime ($d\ll X$), the light intensity at the detector on a circular screen is given by~\cite{BORN64}
\begin{equation}
I(\theta) = A\left(\frac{\sin\frac{qa\sin\theta}{2}}{\frac{qa\sin\theta}{2}}\right)^2 \cos^2\frac{qd\sin\theta}{2},
\label{tbi2}
\end{equation}
where $A$ is a constant, $q=2\pi f/c$
and $\theta$ denotes the angular position of the detector $D$ on the circular screen, see Fig.~\ref{exptbi}.

\begin{figure}[t]
\begin{center}
\includegraphics[width=8cm]{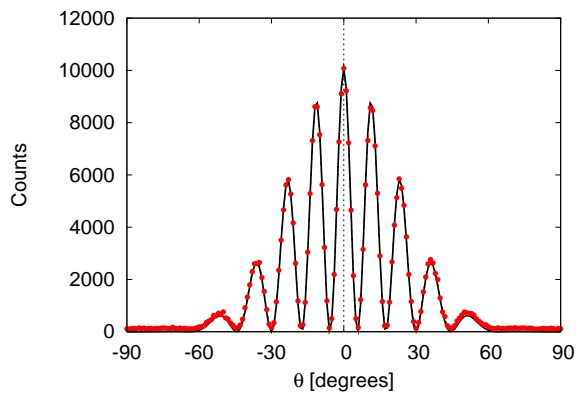}
\caption{%
Detector counts (circles) as a function of the angular detector position $\theta$
as obtained from the EBCM simulation of the two-beam interference experiment depicted in Fig.~\ref{exptbi}.
The solid line is a least-square fit of the EBCM data to the prediction of wave theory, Eq.~(\ref{tbi2}).
The sources have a width $a=c/f$ and are separated by a distance $d=5c/f$
and emit particles according to the current distribution Eq.~(\ref{tbi1}).
The source-detector distance $X=100c/f$, see Fig.~\ref{exptbi}.
Simulation parameters: On average, each of the 181 detectors receives $10^4$ messengers,
$\gamma=\widehat\gamma=0.99$ and $N_p=500$.
}
\label{simtbi}
\end{center}
\end{figure}

\subsubsection{EBCM simulation}

If it is true that individual particles build up the interference pattern one by one
and that there is no direct communication between the particles,
simply looking at Fig.~\ref{exptbi} leads to the logically unescapable
conclusion that the interference pattern can only be due to the internal operation of the detector~\cite{PFlE67}:
There is nothing else that can cause the interference pattern to appear.
Of course, the EBCM of a detector is designed to cope with this task.

In the EBCM, the messengers leave the source one-by-one, at positions $y$ drawn randomly from
a uniform distribution over the interval $[-d/2-a/2,-d/2+a/2]\cup[+d/2-a/2,+d/2+a/2]$, see Eq.~(\ref{tbi1}).
The angle $\beta$ is a uniform pseudo-random number between $-\pi/2$ and $\pi/2$.
When a messenger is created, its time-of-flight is set to zero.
For simplicity, we consider the case of fully polarized light (the same $\xi$ for all messagers) only.

When a messenger travels from the source at $(0,y)$ to the circular detector screen with radius $X$,
it updates its own time-of-flight.
Specifically, a messenger leaving the source at $(0, y)$ under an angle $\beta$ (see Fig.~\ref{exptbi})
will hit the detector screen at a position determined by the angle $\theta$ given by
\begin{equation}
\sin\theta=\frac{y\cos^2\beta+\sin\beta\sqrt{X^2-y^2\cos^2\beta}}{X}
,
\end{equation}
where $|y/X|< 1$.
The time-of-flight $t$ is then given by
\begin{equation}
t=\frac{\sqrt{X^2-2yX\sin\theta+y^2}}{c}
.
\end{equation}
As a messenger hits a detector, this detector updates its internal state (the internal states of all other detectors
do not change) using the data contained in the message and then decides whether to generate a zero or a one output
event.
Only after the messenger has been processed by the detector, the source is allowed to emit a new messenger.
This process is then repeated many times.

In Fig.~\ref{simtbi}, we present simulation results for a representative case.
A Mathematica implementation that uses a variant~\cite{JIN09a} of the EBCM used in this paper
can be downloaded from the Wolfram Demonstration Project web site~\cite{DS08}.

From Fig.~\ref{simtbi} it is clear that the EBCM reproduces the results of \MT\
without taking recourse of the solution of a wave equation
and that the detection efficiency of the detectors is very close to 100\%.
In other words, the interference patterns generated by EBCM cannot be attributed
to ineffecient detectors.

\subsection{Remarks}

It is of interest to compare the detection counts, observed in the EBCM simulation
of the two-beam interference experiment, with those observed in a real experiment with single photons~\cite{JACQ05}.
In the simulation that yields the results of Fig.~\ref{simtbi}, each of the 181 detectors making up the
detection area is hit on average by ten thousand photons and the total number number of clicks generated
by the detectors is 296444. Hence, the ratio of the total number of detected to emitted photons is
of the order of 0.16, orders of magnitude larger than
the ratio $0.5\times 10^{-3}$ observed in the single-photon interference experiments~\cite{JACQ05}.

\section{Optical Components}\label{optical}

The EBCM of optical components such as wave plates and beam splitters
can be constructed by connecting several units that simulate the interfaces (with suitable parameters)
of these components~\cite{BINH10}.
From a simulation point of view, such a construction is both inefficient and unnecessary.
Having shown that two connected EBCM's of an interface reproduce the result
of a plane parallel plate, it is legitimate to replace the two interfaces
by one ``lumped'' EBCM that simulates a beam splitter for instance~\cite{BINH10}.
In this section, we specificy the lumped EBCM of optical components
that are used in quantum optics experiments.

\subsection{Beam splitter}

The processing unit that acts as a beam splitter (BS)
has the generic structure, depicted in Fig.~\ref{figmachine0}.
The device has two input and two output ports labeled by $k=0,1$
and consists of an input stage, a DLM,
a transformation stage (T), and an output stage (O)~\cite{RAED05d,RAED05b,RAED05c,MICH05}.
In fact, up to the transformation matrix $\mathbf{T}$, the processing unit
is identical to the one of a single interface.

For a 50-50 BS, the transformation matrix $\mathbf{T}$ reads
\begin{equation}
\mathbf{T}=
\frac{1}{\sqrt{2}}
\left(\begin{array}{lccr}
        1&i&0&0\\
        i&1&0&0\\
        0&0&1&i\\
        0&0&i&1
\end{array}\right)
,
\label{opti0}
\end{equation}
which is the same as the one employed in wave theory.

\subsection{Polarizing Beam Splitter}

A polarizing beam splitter (PBS) is used to redirect the photons depending on their polarization.
For simplicity, we assume that the coordinate system used to define
the incoming messages coincides with the coordinate system defined
by two orthogonal directions of polarization of the PBS.

The processing unit that acts as a PBS has the generic structure depicted in Fig.~\ref{figmachine0}.
The unit differs from the previous ones in the details of the transformation matrix $\mathbf{T}$ only.
For instance, if the PBS passes $S$-polarized light, it reflects $P$-polarized light.
In this case, the transformation matrix $\mathbf{T}$ of a PBS reads
\begin{equation}
\mathbf{T}=
\left(\begin{array}{lccr}
        1&0&0&0\\
        0&1&0&0\\
        0&0&0&i\\
        0&0&i&0
\end{array}\right)
,
\label{opti1}
\end{equation}
which is the same as the one employed in wave theory.

\subsection{Wave plates}

\begin{figure}[t]
\begin{center}
\includegraphics[width=5cm]{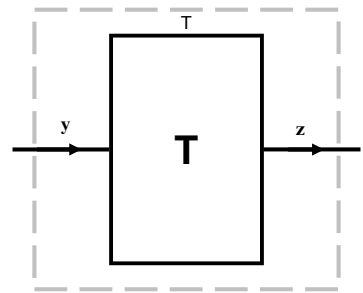}
\caption{Diagram of a processing unit that performs an event-based simulation of wave plates.
Compared to the generic diagram Fig.~\ref{figmachine0}, the DLM stage and the output stage are missing.
The solid lines represent the input and output ports of the device.
The presence of a message is indicated by an arrow on the corresponding port line.
The transformation matrices ${\bf T}$ for the HWP and QWP are
given by Eq.~(\ref{opti2}) and Eq.~(\ref{opti3}), respectively.
}
\label{figwaveplate}
\end{center}
\end{figure}

A half-wave plate (HWP) changes the polarization of the light but also changes its phase.
A quarter-wave plate (QWP) changes the polarization of the light and induces
a phase difference between the $S$ and $P$ components.
In optics, both plates are often used as retarders.
In the EBCM, this retardation of the wave corresponds to a change in the time-of-flight of the messenger.

In contrast to the BS and PBS, as lumped devices
the HWP and QWP may be simulated without the input stage, that is the DLM may be omitted.
The lumped device has only one input and one output port, as shown in Fig.~\ref{figwaveplate},
and we therefore omit the subscript that labels that input and output port in the following.

The transformation performed by the HWP reads
\begin{equation}
\mathbf{T}
=
-i
\left(\begin{array}{lr}
        \cos2\theta&\sin2\theta\\
        \sin2\theta&-\cos2\theta
\end{array}\right)
,
\label{opti2}
\end{equation}
where $\theta$ denotes the angle of the optical axis with respect to the laboratory frame.
Similarly, the transformation performed by the QWP is
\begin{equation}
\mathbf{T}
=
\frac{1}{\sqrt{2}}
\left(\begin{array}{lr}
        1-i\cos2\theta&-i\sin2\theta\\
        -i\sin2\theta&1+i\cos2\theta
\end{array}\right)
.
\label{opti3}
\end{equation}

\subsection{Ideal mirror}

The EBCM model of an interface with $\widetilde r_S=\widetilde r_P=1$ can be used
to simulate the operation of an ideal mirror but from a computational viewpoint,
it is both inefficient and unnecessary to simulate the ideal mirror in this manner.
An ideal mirror merely acts as a hard wall at which the messengers
undergo an elastic scattering event and the second component of the message $\mathbf{y}$ changes sign.
In practice, these rules are trivial to implement in a simulation.

\section{Single-photon quantum optics experiments}\label{sec6}\label{applications}

In the EBCM approach, the processing units that simulate the optical components are connected in such a way that the simulation setup
is an exact one-to-one copy of the laboratory experiment.
The source sends messengers one-by-one but at all times, there is at most one messengers being routed through the network of
processing units. Only after a detector has processed the messenger, the source is allowed to create a new messenger.
This procedure guarantuees that the simulation process trivially satisfies Einstein's criterion of local causality.
Needless to say, the internal state of the processors may be given an interpretation in terms of the polarization,
displacement or any other classical concept that describes the state of the material (atoms, molecules...).
In other words, all the elements of the EBCM may be given an interpretation within the realm of classical physics.
As any such interpretation, which would be necessarily subjective, has no impact on the facts produced by the EBCM, we will not
engage in the endeavour to give such an interpretation.

\subsection{Mach-Zehnder interferometer}\label{MZI}

\begin{figure}[t]
\begin{center}
\includegraphics[width=8cm]{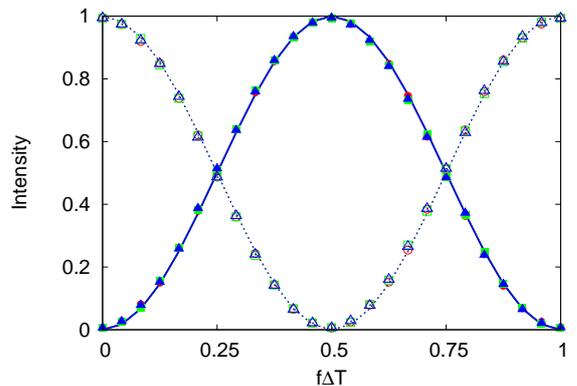}
\caption{%
EBCM simulation results for the number of particles
registered by detectors $D_0$ (solid markers) and $D_1$ (open markers), divided by the total count of detected particles,
as a function of the difference $\Delta T=(T_0-T_1(x))$ between the time-of-flight in the lower and upper arm
of the MZI, respectively.
The solid and dashed lines are the exact results of \MT\ for the normalized intensities at $D_0$ and $D_1$,
respectively.
Simulation parameters: $10^4$ events per pair of open en closed markers, $\gamma=\widehat\gamma=0.99$.
The markers for different polarizations (for the legend, see Fig.~\ref{simplate1}) coincide, in agreement with \MT.
}%
\label{simmzi}
\end{center}
\end{figure}

A description of the MZI was given in Section~\ref{comput}.
In Fig.~\ref{simmzi}, we present the EBCM simulation results for the number of particles
registered by detectors $D_0$ and $D_1$, divided by the total count of detected particles,
as a function of the difference between the time-of-flight in the lower and upper arm
of the MZI, respectively.
In this simulation, we set $T_1(x)=T_0-2\pi x/f$ such that $f\Delta T=f(T_0-T_1(x))=2\pi x$
and vary $x$ from zero to one.
From Fig.~\ref{simmzi}, it is clear that the EBCM model reproduces the results of
\MT.

The MZI may well be the simplest example that
illustrates the conceptual difficulties that arise when one tries
to apply concepts of  \QT\ to give a rational, logically consistent, explanation of what actually happens
to the photons in the MZI.
As shown in the experiment of Grangier {\sl et al}~\cite{GRAN86},
photons may be considered as indivisable particles~\cite{GARR09},
each one following one particular path through the MZI.
Assuming that the observed interference in this experiment can only be a wave phenomenon,
the somewhat mystical concept of wave-particle duality is introduced.
It is said that photons exhibit both wave and particle behavior depending upon the circumstances of
the experiment~\cite{HOME97}, in contradiction with the experiment~\cite{GRAN86} which shows that
photons are indivisable particles.
As a last resort to save the quantum theoretical explanation, another mystical element, namely the collapse of the wave function is introduced.
The collapse mechanism has remained mystical after hunderd years of using  \QT: A logically and physically acceptable,
experimentally testable mechanism has not been found.
In effect, the main purpose of these two mystical elements is to demonstrate, what is long known,
that  \QT\ does not provide any insight into what happens in the system.
It does not contain the elements to give a cause-and-effect description
of natural phenomena but describes the statistical properties of our observations very well.
Instead of searching for rational cause-and-effect descriptions,
to get out of the logical mess created by introducing elements of magic,  \QT\ simply
postulates that such a rational cause-and-effect description does not exist.

In the EBCM approach, there is no need to invoke elements of magic or irrational reasoning
to explain the results of the MZI experiment of Grangier {\sl et al}~\cite{GRAN86}.
The crux is that realize that interference is not necessarily a wave phenomenon
but can be generated by particles that interact with some agent (the processors in the
EBCM).

\subsection{Wheeler's delayed choice experiment}\label{WDC}

In 1978, Wheeler proposed a gedanken
experiment~\cite{WHEE83}, a variation on Young's double slit experiment, in which the
decision to observe wave or particle behavior is made after the photon has
passed the slits. The pictorial description of this experiment defies common sense:
The behavior of the photon in the past is said to be changing from a particle to a wave or vice versa.

\subsubsection{Experiment}

\begin{figure}[t]
\begin{center}
\includegraphics[width=8cm]{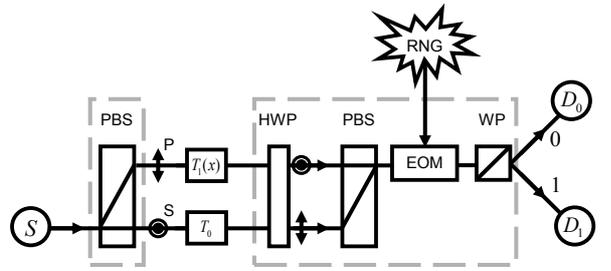}
\caption{%
Schematic diagram of the experimental setup for Wheeler's delayed-choice
experiment with single photons~\cite{JACQ07}.
EOM: electro-optic modulator; RNG: Random number generator;
WP: Wollaston prism (= PBS); $D_{0}$, $D_{1}$: Detectors.
The EBCM replaces the physical devices by the corresponding message-processing units.
}%
\label{expwdc}
\end{center}
\end{figure}

In an experimental realization of Wheeler's delayed choice experiment,
Jacques {\sl et al.}~\cite{JACQ07} send linearly polarized single photons
through a polarizing beam splitter (PBS) that
together with a second, movable PBS forms an interferometer (see Fig.~\ref{expwdc}).
Moving the second PBS induces a time-delay in one of the arms of the interferometer~\cite{JACQ07},
symbolically represented by $T_1(x)$ in Fig.~\ref{expwdc}.
The electro-optic modelator (EOM) performs the same function as a HWP.
Changing the electrical potential applied to the EOM, changes the rotational angle that mixes the two polarizations,
see Eq.~(\ref{opti2}).
In the experiment~\cite{JACQ07}, the random number generator RNG generates a sequence of binary random numbers $r_n=0,1$.
The output of RNG is used to control the potential applied to the EOM.
If $r_n=0$, the rotation angle $\theta_{\mathrm{EOM}}=0$ and if $r_1$, $\theta_{\mathrm{EOM}}=\pi/8$.
In the former case, there can be no interference: There is no mixing of $S$- and $P$-polarized light.
In the latter case, interference can occur and we expect an interference pattern
that is the same as the one of a MZI.

Detector $D_0$ ($D_1$) counts the events generated at port 0 (1) of the Wollaston prism (WP).
During a run of $N$ events, the algorithm generates the data set
\begin{equation}
\Gamma =\left\{d_{n},r_{n}|n=1,...,N;f\Delta T\right\}
,
\end{equation}
where $d_{n}=0$ ($1$) indicates that detector $D_{0}$ ($D_{1}$) fired
and $r_{n}=0,1$ is a binary pseudo-random number that is chosen
after the $n$th message has passed the first PBS.

\subsubsection{EBCM simulation}
\begin{figure}[t]
\begin{center}
\includegraphics[width=8cm]{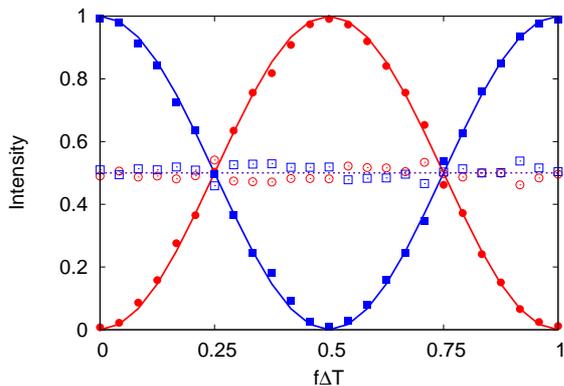}
\caption{%
EBCM simulation results (markers) for the Wheeler delayed-choice experiment depicted in Fig.~\ref{expwdc}.
The number of particles registered by detectors $D_0$ and $D_1$
divided by the total count of detected particles,
as a function of the difference $\Delta T=(T_0-T_1(x))$ between the time-of-flight in the lower and upper arm
of the interferometer, respectively.
Open circles: Intensity measured by $D_0$ for $\theta_{\mathrm{EOM}}=0$ ($r_n=0$);
Open squares: Intensity measured by $D_1$ for $\theta_{\mathrm{EOM}}=0$;
Closed circles: Intensity measured by $D_0$ for $\theta_{\mathrm{EOM}}=\pi/4$ ($r_n=1$);
Closed squares: Intensity measured by $D_1$ for $\theta_{\mathrm{EOM}}=\pi/4$;
Dashed line: Intensities measured by $D_0$ and $D_1$ as obtained from \MT\ for $\theta_{\mathrm{EOM}}=0$;
Solid red (blue) line: Intensity measured by $D_0$ ($D_1$) as obtained from \MT\ for $\theta_{\mathrm{EOM}}=\pi/4$.
Simulation parameters: $2600$ emitted events per pair of open/closed markers,
$\gamma=\widehat\gamma=0.99$ and $N_p=1$.
The actual counts ($= 1300\times\mathrm{intensities}$) are in good quantitative agreement with
the experimental results reported in Ref.~\cite{JACQ07}.
}%
\label{simwdc}
\end{center}
\end{figure}

We simulate the Wheeler delayed-choice experiment of Fig.~\ref{expwdc} by connecting the various EBCM of the
optical components in exactly the same manner as in Fig.~\ref{expwdc}.
The simulation generates the data set $\Gamma$ just like in the experiment~\cite{JACQ08,JACQ07}
and is analyzed in the same manner.
The EBCM simulation results presented in Fig.~\ref{simwdc} show that there is excellent
agreement between the event-based simulation data and the predictions of wave theory.

Of course, in the case of the classical, locally causal EBCM, there is no need
to resort to concepts such as particle-wave duality and the mysteries of delayed-choice
to give a rational explanation of the observed phenomena.
In the simulation we can always track the particles, independent of $r_n=0,1$.
These particals always have full which-way information,
never directly communicate with each other, arrive one by one at a detector
but nevertheless build up an interference pattern at the detector if $r_n=1$.
Thus, the EBCM of Wheeler's delayed-choice experiment provides a unified
particle-only description of both cases $r_n=0,1$ that does not defy common sense.

\subsection{Quantum Eraser}\label{QE}
In 1982, Scully and Dr{\"{u}}hl proposed a photon interference experiment,
called ``quantum eraser''~\cite{SCUL82}, in which
the photons are labelled by which-way markers.
In this experiment, the which-way information of the photons is known to the
experimenter and hence, according to common lore, no interference is to be expected.
However by erasing the which-way information afterwards by a ``quantum eraser'',
the interference pattern can be recovered~\cite{SCUL82}, even from data that
has been recorded and saved in a file~\cite{KIM00}.

Quantum eraser experiments have been described ``as one of the most intriguing effects in quantum mechanics'',
but have also been regarded as ``the fallacy of delayed choice and quantum eraser''~\cite{AHAR05}.
Clearly, they challenge the point of view
that the wave and particle behavior of photons are complementary:
The observation of interference, commonly associated with wave behavior,
depends on the way the data is analyzed after the photons
have passed through the interferometer.

\begin{figure}[t]
\begin{center}
\includegraphics[width=8cm]{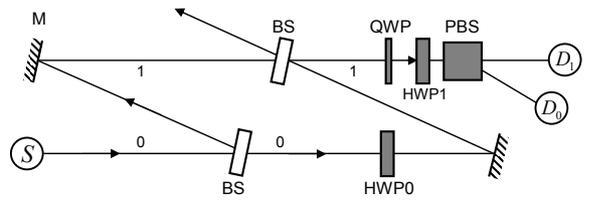}
\caption{%
Schematic diagram of the experimental setup for the quantum eraser experiment
with photons studied in Ref.~\cite{SCHW99}.
Moving the mirror labeled M changes the difference $\Delta T$ between the time-of-flight in the lower and upper arm
of the interferometer.
}%
\label{expqe}
\end{center}
\end{figure}

The quantum eraser has been implemented in several different experiments
with photons, atoms, etc.~\cite{SCUL91,KWIA92,PITT96,SCHW99,KIM00,WALB02,SCAR07}.
In this paper, we consider the experiment performed by Schwindt {\sl et al.}~\cite{SCHW99}
in which the polarization of the photons is employed to encode the which-way information.
The experimental setup (see Fig.~\ref{expqe}) consists of a linearly polarized
single-photon source (not shown), a MZI
of which the time-of-flight of path 1 (see Fig.~\ref{expqe}) can be varied by moving the mirror $M$
and an analysis system which is a combination of a QWP,
a HWP (HWP1), and a calcite prism (WP) operating as a PBS.
Another adjustable HWP (HWP0) is inserted in path 0 of the MZI to entangle the photon's path with its polarization~\cite{SCHW99}.

\begin{figure}[t]
\begin{center}
\includegraphics[width=8cm]{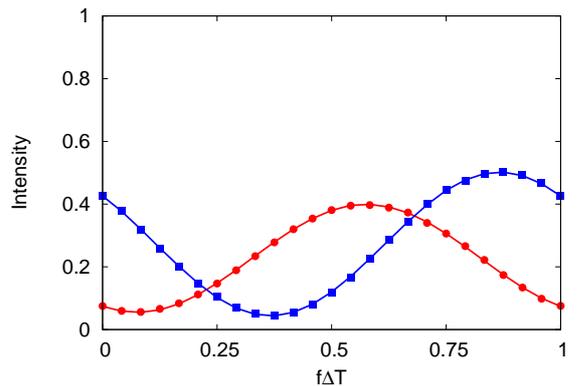}
\caption{%
The number of particles registered by detectors $D_0$ and $D_1$
divided by the total count of detected particles as a function of the time-of-flight difference $\Delta T$ for
$\theta_0=\pi/3$, $\theta_1=\pi/4$, and $\theta_2=\pi/8$,
as obtained from the EBCM simulation of the quantum eraser experiment depicted in Fig.~\ref{expqe}.
Circles: Intensity measured by $D_0$;
Squares: Intensity measured by $D_1$;
Solid red (blue) line: Intensity measured by $D_0$ ($D_1$) as obtained from \MT, see Eq.~(\ref{qe0}).
Simulation parameters: $10^4$ events, $\gamma=\widehat\gamma=0.99$ and $N_p=1$.
}%
\label{simqe}
\end{center}
\end{figure}

\subsubsection{Wave theory}
If a photon, described by a pure $S$-polarized state is injected into the interferometer with the HWP0 set to $45^\circ$,
the photon that arrives at the second BS of the MZI carries a which-way marker:
The photon will be in the $P$ polarized state if it followed path 0 and will be in the $S$ polarized state if it followed path 1.
If the rotation angle of HWP1 is zero, there will be no interference
and the detectors reveal the full which-way information of each detected photon.
If the rotation angle of HWP1 is nonzero, the $S$- and $P$-polarized states interfere,
the which-way information of each photon will be partially or completely ``erased''.
Thus, by varying the rotation angle of HWP1, the illusion is created
that the character of the photon in the MZI ``changes'' from particle to wave and vice versa.

Assuming that the photons emitted by the source are described by a pure $S$-polarized state,
wave theory predicts that the intensities at the detectors $D_0$ and $D_1$ are given by
\begin{eqnarray}
I_0&=&\frac{1}{16} \left\{ 4-\cos 4(\theta_2-\theta_1) -\cos 4\theta_1\right.
\nonumber \\
&& \left. -\cos 4(\theta_2-\theta_1-\theta_0)-\cos4(\theta_1-\theta_0) \right.
\nonumber \\
&& \left. +4 \cos f\Delta T \sin (2\theta_2-4\theta_1) \sin 2\theta_0  \right.
\nonumber \\
&& \left.  -2 \sin f\Delta T [ \cos (4\theta_2-4\theta_1-2\theta_0) \right.
\nonumber \\
&& \left.
+\cos (4\theta_1-2\theta_0) -2 \cos 2\theta_0 ]
\right\}
\\
I_1&=&\frac{1}{16} \left\{ 4+\cos 4(\theta_2-\theta_1) +\cos 4\theta_2\right.
\nonumber \\
&& \left.+\cos 4(\theta_2-\theta_1-\theta_0)  +\cos4(\theta_1-\theta_0) \right.
\nonumber \\
&& \left.-4 \cos f\Delta T \sin (2\theta_2-4\theta_1) \sin 2\theta_0 \right.
\nonumber \\
&& \left.
 +2 \sin f\Delta T [ \cos (4\theta_2-4\theta_1-2\theta_0) \right.
\nonumber \\
&& \left.
+\cos (4\theta_1-2\theta_0)+2 \cos 2\theta_0 ]
\right\}
,
\label{qe0}
\end{eqnarray}
where $\theta_0$, $\theta_1$, and $\theta_2$ are the rotation angles of HWP0, HWP1, and QWP, respectively.

\subsubsection{EBCM simulation}
In an earlier paper in this journal, it was shown that quantum eraser experiments can be simulated
without invoking concepts of quantum theory and without first solving a wave mechanical problem~\cite{JIN09c}.
In this paper, we demonstrate that the same experiment can be simulated by simply re-using the EBCM of the optical components.

In Fig.~\ref{simqe} we present the results of an event-by-event simulation
of the experiment depicted in Fig.~\ref{expqe}.
It is clear that there is excellent agreement between the EBCM data and the prediction of  \QT\
for the system described by the pure state.
Elsewhere, we have already demonstrated that the EBCM can also cope with
systems that  \QT\ describes in terms of mixed states~\cite{JIN09c}.
The EBCM correctly describes the outcome of the quantum
eraser experiment of Fig.~\ref{expqe} under all circumstances~\cite{JIN09c}.

\subsection{Single-photon tunneling}\label{TUN}

\begin{figure}[t]
\begin{center}
\includegraphics[width=8cm]{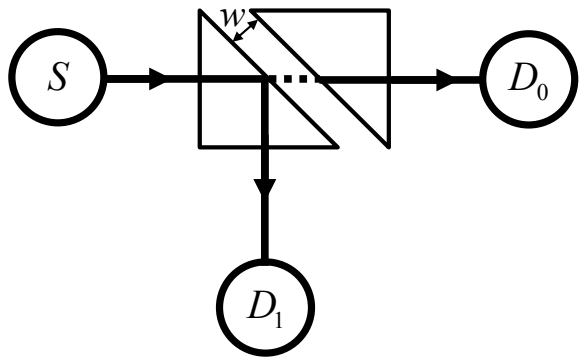} 
\caption{%
Schematic diagram of the experimental setup for the photon tunneling experiment Ref.~\cite{MIZO92}.
The hypotenuses of the identical prisms are separated from each other by a distance $w$.
}%
\label{exptun}
\end{center}
\end{figure}

A conceptually simple, direct experimental demonstration that photons behave as
indivisable entities and that displays both particle and wave characteristics
was proposed by Ghose et al.~\cite{GHOS91}, drawing inspiration from an experiment
that was carried out with microwaves almost hundred years earlier~\cite{BOSE1897}.
The experimental setup is sketched in Fig.~\ref{exptun}.

It is instructive to consider \MT\ for this experiment first.
A light wave is sent to a prism that is separated from another prism by a distance $w$.
If $w$ is very large, the total internal reflection in the first prism causes virtually
all light to be reflected~\cite{BORN64}.
If $w$ is of the order of the wavelength or less, evanescent waves at the surface
of the first prism can penetrate the second prism, giving rise to a measurable
flux of light in the second prism~\cite{BOSE1897}.
It is said that the wave can ``tunnel'' through the gap separating the prisms.
In the case of waves, the experimental observation of this tunneling effect can
be completely understood within the framework of \MT.

Let us now try to explain this observation if we accept, as we do in this paper,
that light is made out of indivisable entities, that is out of photons.
Assuming ideal single-photon detectors and a single-photon source,
the experiment may show that
\begin{enumerate}
\item{Both detectors $D_0$ and $D_1$ click within a time window, much smaller
than the time between two successive single-photon emissions.}
\item{Detectors $D_0$ and $D_1$ click in perfect anti-coincidence.}
\end{enumerate}
Experimental results favor possibility (2)~\cite{MIZO92,UNNI96,BRID04} (see also our
discussion of the experiment by Grangier {\sl et al.} in Section~\ref{pie})
and within the standard interpretation of  \QT\ are considered to be a proof
that the particle-wave duality must be considered in weak sense
and not in the orginal ``complementarity'' sense of Bohr~\cite{HOME97,BRID04}.
There is not much point to dwell on this issue here because,
as we show now, the EBCM approach provides a simple picture of
this phenomenon without introducing the mystique of particle-wave duality.

\begin{figure}[t]
\begin{center}
\includegraphics[width=8cm]{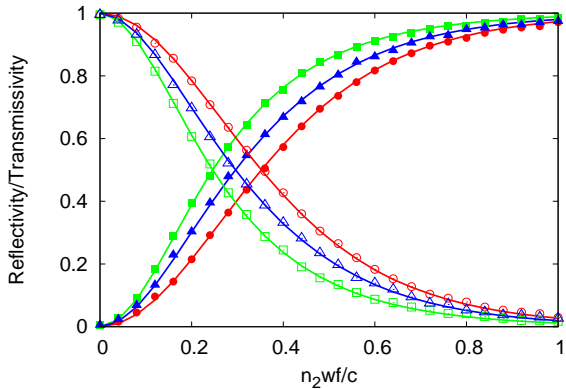}
\caption{%
The number of particles registered by detectors $D_0$ (transmissivity, open markers) and $D_1$ (reflectivity, closed markers)
divided by the total count of detected particles as a function of the
width $w$ of the gap between the two prisms ($n=1.52$),
as obtained from an EBCM simulation of the photon tunneling experiment depicted in Fig.~\ref{exptun}.
The solid lines are the exact results of \MT~\cite{BORN64}.
Green markers and lines: $\xi=0$ ($S$-polarization);
Blue markers and lines: $\xi=\pi/4$ ($S$+$P$ polarization);
Red markers and lines:  $\xi=\pi/2$ ($P$-polarization).
Simulation parameters: $10^4$ events per marker, $\gamma=\widehat\gamma=0.99$ and $N_p=1$.
}%
\label{simtun}
\end{center}
\end{figure}

The results of an EBCM simulation, using the ``lumped device'' representation
of the two-prism system, of the experiment depicted in Fig.~\ref{exptun}
are shown in Fig.~\ref{simtun}, together with the predictions of \MT.
It is clear that there is excellent agreement between the corpuscular
model and the wave model, demonstrating once more that the
predictions of the latter can be obtained from a simulation that
involves particles only.


\section{Photon correlation experiments}\label{PCE}

\subsection{Einstein-Podolsky-Rosen-Bohm experiment}\label{EPRB}

To construct an event-based computer simulation model of the EPRB
experiment with photons performed by Weihs et al.~\cite{WEIH98}),
we use the same EBCMs for the optical components as in the previous sections.
The resulting EBCM for the whole EPRB experiment is substantially different from
our earlier event-based simulation models that reproduce the result of \QT\ for the singlet state
and product state~\cite{RAED06c,RAED07a,RAED07b,RAED07c,RAED07d,ZHAO08b}.
The difference is that in our earlier work we adpoted the tradition
in this particular subfield to use simplified mathematical models for the optical components
that, when re-used for different optics experimements, would fail to reproduce the results
of these experiments. In this section, we simply use (without any modification) the
EBCMs for the optical components that are present in the laboratory experiment
and analyze the simulation data in exactly the same manner as the experimental
data for this experiment has been analyzed~\cite{WEIH98}.

\subsubsection{Laboratory experiment}
\begin{figure}[t]
\begin{center}
\includegraphics[width=8cm]{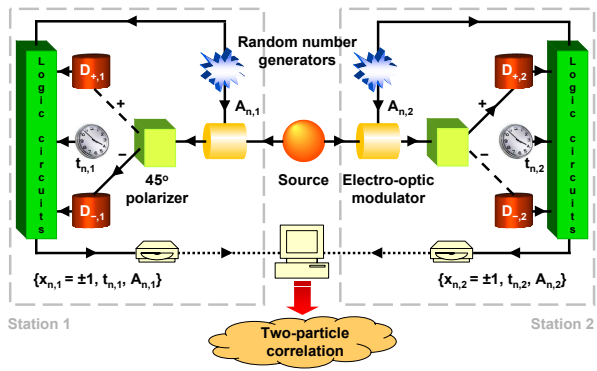}
\caption{Schematic diagram of an EPRB experiment with photons~\cite{WEIH98}.
}
\label{expeprb}
\end{center}
\end{figure}

In Fig.~\ref{expeprb}, we show a schematic diagram of an EPRB experiment with photons (see also Fig.~2 in~\cite{WEIH98}).
The source emits pairs of photons.
Each photon of a pair travels to an observation station in which it is manipulated and detected.
The two stations are assumed to be identical.
They are separated spatially and temporally, preventing the observation at station 1 (2) to have a causal effect on the
data registered at station $2$ (1)~\cite{WEIH98}.
As the photon arrives at station $i=1,2$, it passes through an EOM 
that rotates the polarization of the photon by an angle depending on the voltage applied to the modulator.
These voltages are controlled by two independent binary random number generators.
A PBS sends the photon to one of the two detectors.
The station's clock assigns a time-tag to each generated signal.
We consider two different experiments, one in which the source emits photons
with opposite but otherwise unpredictable polarization and those with a source emitting
photons with fixed polarization.

\subsubsection{Quantum theory}

The quantum theoretical description of the EPRB experiment with photons
exploits the fact that the two-dimensional vector space spanned by two orthogonal polarization vectors
is isomorphic to the vector space of spin-1/2 particles.
The predictions of \QT\  for the single and two-particle averages
for an experiment described by a quantum system of two spin-1/2 particles in the singlet state and
a product state are given in Table~\ref{tab2} and serve as a reference for the EBCM simulation.

\begin{table}
\begin{center}
\caption{%
The single spin expectations, the two-spin expectation and the spin-spin correlation of
two spin-1/2 particles in the singlet state
$|\mathrm{Singlet}\rangle=\left( {| {SP} \rangle -| {PS} \rangle } \right)/\sqrt{2} $
and in the product state
$|\mathrm{Product}\rangle=(\cos \eta_1|P\rangle_1 +\sin \eta_1|S\rangle_1) (\cos \eta_2 |P\rangle_2 +\sin \eta_2 |S\rangle _2)$.
}
\label{tab2}       
\begin{ruledtabular}
\begin{tabular}{lcc}
& $|\mathrm{Singlet}\rangle$  & $|\mathrm{Product}\rangle$  \\
\noalign{\smallskip}\hline\noalign{\smallskip}
$\widehat E_1(\alpha_1)$         & $0$ & $\cos2(\alpha_1-\eta_1)$\\
$\widehat E_2(\alpha_2)$         & $0$ & $\cos2(\alpha_2-\eta_2)$\\
$\widehat E_{12}(\alpha_1,\alpha_2)$ & $-\cos2(\alpha_1-\alpha_2)$ & $\cos 2(\alpha_1-\eta_1)\cos 2(\alpha_2-\eta_2)$\\
$\widehat \rho_{12}(\alpha_1,\alpha_2)$ & $\widehat E_{12}(\alpha_1,\alpha_2)$ & 0
\end{tabular}
\end{ruledtabular}
\end{center}
\end{table}

\subsubsection{EBCM of the EPRB experiment}
The detectors, PBSs and EOMs are simulated using their EBCMs described earlier.
Recall that the EBCM of the detector mimics a detector with 100\%
detection efficiency.
The EBCM of the EPRB experiment trivially satisfies Einstein's criteria of local causality,
does not rely on any concept of quantum theory and, as will be shown below,
reproduces the results of quantum theory for both types of experiments.

\begin{figure}[t]
\begin{center}
\includegraphics[width=8cm]{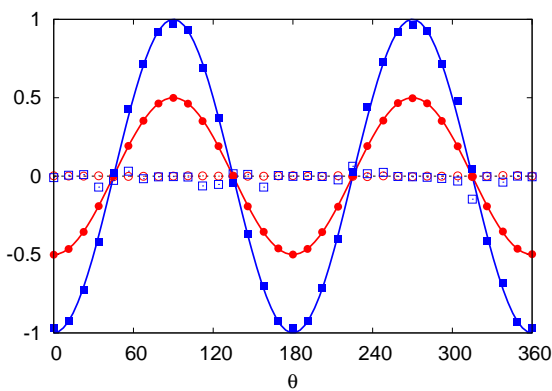}
\caption{%
Computer simulation data and quantum theoretical results
for the single-particle and two-particle averages
of the EPRB experiment depicted in Fig.~\ref{expeprb}
as a function of $\theta=\alpha_1-\alpha_2$.
The source emits particles with opposite polarization,
uniformly distributed over the unit circle.
Open circles and squares: Eq.~(\ref{eprb3b}) and \ref{eprb3b}), respectively;
Closed squares: Eq.~(\ref{eprb5a})
Closed circles: Eq.~(\ref{eprb5b}) for $W=T_{\mathrm{EPRB}}$,  that is
by ignoring the time-tag data;
Blue solid line: $\widehat E_{12}(\alpha_1,\alpha_2)=-\cos2\theta$;
Red solid line: $-2^{-1}\cos2\theta$.
Simulation parameters: $3\times10^5$ pairs, $T_{\mathbf{EPRB}}=1000$, $W=1$, $d=4$, $\gamma=\widehat\gamma=0.99$ and $N_p=1$.
Note that the total number of pairs emitted by the source is about
the same as the number of photons per station detected in the experiment
reported in Ref.~\cite{WEIH98} (experimental data set called longdist35).
}
\label{simeprb1}
\end{center}
\end{figure}

\begin{figure}[t]
\begin{center}
\includegraphics[width=8cm]{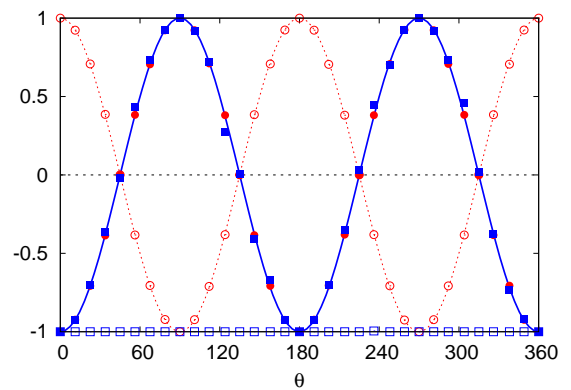}
\caption{%
Computer simulation data and quantum theoretical results
for the single-particle and two-particle averages
of the EPRB experiment depicted in Fig.~\ref{expeprb}
as a function of $\theta=\alpha_1-\eta_1$ with $\alpha_2=0$.
Same as in Fig.~\ref{simeprb1} except that the source emits particles with
with $P$ ($\eta_1=0$) and $S$ ($\eta_2=\pi/2$) polarization only.
Open circles and squares: Eq.~(\ref{eprb3b}) and \ref{eprb3b}), respectively;
Closed squares: Eq.~(\ref{eprb5a})
Closed circles: Eq.~(\ref{eprb5b}) for $W=T_{\mathrm{EPRB}}$,  that is
by ignoring the time-tag data;
Blue solid line: $\widehat E_{12}(\alpha_1,\alpha_2)=
\cos 2(\alpha_1-\eta_1)\cos 2(\alpha_2-\eta_2)=-\cos2\theta$.
Simulation parameters: Same as in Fig.~\ref{simeprb1}.
}
\label{simeprb2}
\end{center}
\end{figure}

The data generated by the EBCM of the EPRB experiment is analyzed
in exactly the same manner as in experiment~\cite{WEIH98}.
In the simulation, the firing of a detector is regarded as an event.
At the $n$th event, the data recorded on a hard disk at station $i=1,2$
consists of $x_{n,i}=\pm 1$, specifying which of the two detectors fired,
the time tag $t_{n,i}$ indicating the time at which a detector fired,
and the two-dimensional unit vector $\alpha_{n,i}$ that represents the rotation
of the polarization by the EOM at the time of detection.
Hence, the set of data collected at station $i=1,2$ during a run of $N$ events
may be written as
\begin{eqnarray}
\Upsilon_i=\left\{ {x_{n,i} =\pm 1,t_{n,i},\alpha_{n,i} \vert n =1,\ldots ,N } \right\}
\label{eprb1}
.
\end{eqnarray}
In the (computer) experiment, the data $\{\Upsilon_1,\Upsilon_2\}$ may be analyzed
long after the data has been collected~\cite{WEIH98}.
Coincidences are identified by comparing the time differences
$\{ t_{n,1}-t_{n,2} \vert n =1,\ldots ,N \}$ with a time window~\cite{WEIH98} $W$. 
Introducing the symbol $\sum'$ to indicate that the sum
has to be taken over all events that satisfy
$\alpha_i=\alpha_{n,i}$ for $i=1,2$,
for each pair of directions $\alpha_1$ and $\alpha_2$ of the EOMs,
the number of coincidences $C_{xy}\equiv C_{xy}(\alpha_1,\alpha_2)$ between detectors $D_{x,1}$ ($x =\pm 1$) at station
1 and detectors $D_{y,2}$ ($y =\pm1 $) at station 2 is given by
\begin{eqnarray}
C_{xy}&=&\sumprime_{n=1}^N\delta_{x,x_{n ,1}} \delta_{y,x_{n ,2}}
\Theta(W-\vert t_{n,1} -t_{n ,2}\vert)
,
\label{eprb2}
\end{eqnarray}
where $\Theta (t)$ is the Heaviside step function.
We emphasize that we count all events that, according to the same criterion as the one employed in experiment,
correspond to the detection of pairs~\cite{WEIH98}.
The average single-particle counts are defined by
\begin{equation}
E_1(\alpha_1,\alpha_2)=
\frac{\sum_{x,y=\pm1} xC_{xy}}{\sum_{x,y=\pm1} C_{xy}}
,
\label{eprb3a}
\end{equation}
and
\begin{equation}
E_2(\alpha_1,\alpha_2)=\frac{\sum_{x,y=\pm1} yC_{xy}}{\sum_{x,y=\pm1} C_{xy}}
,
\label{eprb3b}
\end{equation}
where the denominator is the sum of all coincidences.

The correlation of two dichotomic variables $x$ and $y$ is defined by~\cite{GRIM95}
\begin{equation}
\rho_{12}(\alpha_1,\alpha_2)=E_{12}(\alpha_1,\alpha_2)-E_1(\alpha_1,\alpha_2)E_2(\alpha_1,\alpha_2)
,
\label{eprb5a}
\end{equation}
where
\begin{eqnarray}
E_{12}(\alpha_1,\alpha_2)&=&\frac{\sum_{x,y} xyC_{xy}}{\sum_{x,y} C_{xy}}
\nonumber \\
&=&\frac{C_{++}+C_{--}-C_{+-}-C_{-+}}{C_{++}+C_{--}+C_{+-}+C_{-+}}
,
\label{eprb5b}
\end{eqnarray}
is the two-particle average.

In general, the values for the average single-particle counts $E_1(\alpha_1,\alpha_2)$ and $E_2(\alpha_1,\alpha_2)$,
the two-particle averages $E(\alpha_1,\alpha_2)$,
and the total number of the coincidences $C(\alpha_1,\alpha_2)=\sum_{x,y=\pm1} C_{xy}(\alpha_1,\alpha_2)$,
not only depend on the directions $\alpha_1$ and $\alpha_2$ but also on
the time window $W$ used to identify the coincidences.

\subsubsection{Time-tag model}

From Eq.~(\ref{eprb1}), it is clear that a model that aims to describe real EPRB experiments
should incorporate a mechanism to produce the time tags $t_{n,1}$ and $t_{n ,2}$.
The importance of incorporating such a mechanism has been pointed out
by S. Pascazio who presented a concrete time-tag model that yields a
good approximation to the correlation of the singlet state (see Table~\ref{tab2})~\cite{PASC86}.

Following our earlier work~\cite{RAED06c,RAED07b,RAED07a,RAED07c,RAED07d},
we assume that as a particle passes through the EOM, it experiences a time delay.
This is a very reasonable assumption as EOMs are in fact used as retarders in optical communication systems.
For simplicity, the time delay ${t}_{n ,i}$ is assumed to be distributed uniformly over the interval $[0,T]$.
From our earlier work we know that the choice $T=T_{\mathbf{EPRB}}\sin^{2d}2(\xi _{n } -\alpha_{n ,i})$
rigorously reproduces the results of Table~\ref{tab2}, that is the results
of quantum theory for the EPRB experiment, if $d=4$ and $W\ll T_{\mathbf{EPRB}}$~\cite{RAED06c,RAED07b,RAED07a,RAED07c,RAED07d}.
Therefore, we adopt this time-tag model in the present paper also.

\subsubsection{Simulation results}
The EBCM generates the data set Eq.~(\ref{eprb1}), just as experiment does~\cite{WEIH98}.
We choose the coincidence window $W$ and compute the coincidences, single-spin and two-spin
averages according to the Eqs.~(\ref{eprb2})--(\ref{eprb5b}).
The simulation results for the two different types of experiments
are presented in Figs.~\ref{simeprb1} and \ref{simeprb2}.
It is obvious that the fully classical, locally causal EBCM reproduces
the results of \QT\  for both the singlet state and the product state.
Note that the detectors have 100\% detection efficiency.

\subsubsection{Remarks}
It is not uncommon to find in the literature,
statements that it is impossible to simulate quantum phenomena by classical processes.
Such statements are thought to be a direct consequence of Bell's theorem~\cite{BELL93} but
are in conflict with other work that has pointed out the irrelevance of Bell's theorem
~\cite{PENA72,FINE74,FINE82,FINE82a,FINE82b,MUYN86,JAYN89,BROD93,FINE96,KHRE99,SICA99,BAER99,%
HESS01,HESS05,ACCA05,KRAC05,SANT05,MORG06,KHRE07,ADEN07,NIEU09,MATZ09,RAED09a}.
This conclusion is supported by several explicit examples that prove
that it is possible to construct algorithms that satisfy
Einstein's criteria for locality and causality, yet reproduce
{\sl exactly} the two-particle correlations of a quantum system in the singlet state,
without invoking any concept of quantum theory~\cite{RAED06c,RAED07a,RAED07b,RAED07c,RAED07d,ZHAO08}.
The simulation results presented in Fig.~\ref{simeprb1}, obtained from different (but similar)
simulation models than the ones used in previous work, provides yet another illustration that
Bell's no-go theorem is of very limited value: It applies to a marginal class
of classical models only and becomes relevant to the EPRB experiments that are performed in the
laboratory if the coincidence window $W$ approaches infinity (on the time scale of the experiment).

\subsection{Hanbury Brown-Twiss experiment}\label{HBT}

HBT experiments~\cite{HBT56a} measure the correlation of light intensities
orginating from two different, uncorrelated sources.
HBT showed that under conditions for which the usual two-beam interference fringes
measured by each of the two detectors vanish, the correlated intensities
of the two detectors can still show interference fringes.
When a HBT experiment is performed with detectors operating in the single-photon-detection regime,
the observation of the fringes in the correlated detector intensities
is attributed to the wave-particle duality of the beam~\cite{GLAU63a,GLAU63b,MAND99}.

\begin{figure}[t]
\begin{center}
\includegraphics[width=8cm]{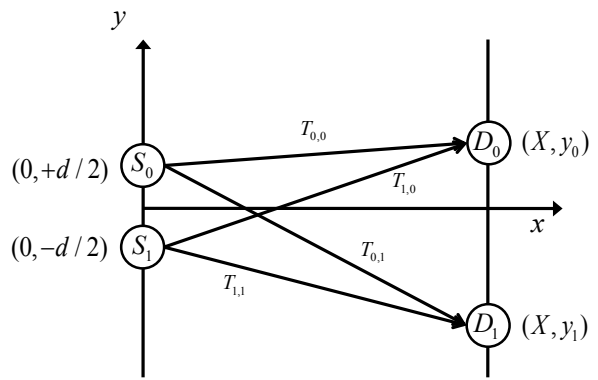}
\caption{Schematic diagram of a HBT experiment.
Single photons emitted from point sources $S_0$ and $S_1$ are registered by two detectors $D_0$ and $D_1$.
The time-of-flight for each of the four possible paths from source $S_m$ to detector $D_n$
is denoted by $T_{m,n}$ where $m,n=0,1$.
}%
\label{exphbt}
\end{center}
\end{figure}

\subsubsection{Scalar wave theory}
Conceptually, the HBT experiment of Fig.~\ref{exphbt} can be viewed as a two-beam experiment with two detectors.
Assume that source $S_m$ ($m=0,1$) emits coherent light of frequency $f$
and produces a wave with amplitude $A_m e^{i\phi_m}$ ($A_m$ and $\phi_m$ real).
For simplicity of presentation, we assume that $A_0=A_1=A$.
According to \MT, the total wave amplitude $B_n$ on detector $n$ is
\begin{equation}
B_n = A \left(e^{i(\phi_0+2\pi fT_{0,n})} + e^{i(\phi_1+2\pi fT_{1,n})}\right)
,
\end{equation}
where the time-of-flight for each of the four possible paths from source $S_m$ to detector $D_n$
is denoted by $T_{m,n}$ where $m,n=0,1$.
The light intensity $I_n = |B_n|^2$ on detector $D_n$ is given by
\begin{equation}
I_n = 2A^2\left\{ 1+ \cos \left[\phi_0-\phi_1+2\pi f(T_{0,n}-T_{1,n})\right]\right\}
.
\label{hbt0}
\end{equation}
If the phase difference $\phi_0-\phi_1$ in Eq.~(\ref{hbt0}) is fixed, the usual two-beam  (first-order) interference fringes are observed.
In this section, $\langle . \rangle$ denotes the average over the variables $\phi_0$ and $\phi_1$.

The essence of the HBT experiment is that if the phase difference $\phi_0-\phi_1$ is a random variable
(uniformly distributed over the interval $[0,2\pi[$) as a function of observation time,
these first-order interference fringes vanish because
\begin{equation}
\langle I_n\rangle = 2A^2.
\label{hbt1}
\end{equation}
However, the average of the product of the intensities is given by
\begin{equation}
\langle I_0I_1\rangle = 4A^4\left( 1+ \frac{1}{2}\cos 2\pi f \Delta T\right)
,
\label{hbt2}
\end{equation}
where $\Delta T =(T_{0,0}-T_{1,0})-(T_{0,1}-T_{1,1})$.
Accordingly, the intensity-intensity correlation Eq.~(\ref{hbt2}) exhibits second-order interference fringes,
a manifestation of the so-called HBT effect.
\begin{figure}[t]
\begin{center}
\includegraphics[width=8cm]{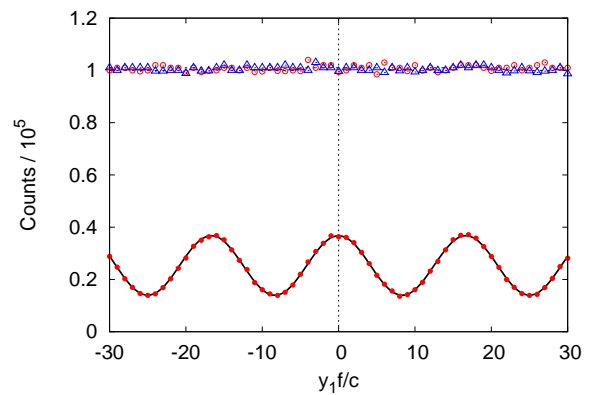}
\caption{%
Computer simulation data of the single-particle and two-particle counts
for the HBT experiment depicted in Fig.~\ref{exphbt}.
Red open circles: EBCM results for the counts of detector $D_0$.
Blue open triangles: EBCM results for the counts of detector $D_1$.
Red closed circles: EBCM results for the coincidence counts.
The dashed and solid lines are least-square fits of the predictions of wave theory
to the EBCM data for the single detector and coincidence counts, respectively.
Simulation parameters: $N_{\mathrm{tot}}=2\times 10^5$ events per $y_1f/c$-value,
$N_{\mathrm{F}}=40$, $X=100000c/f$, $d=2000c/f$, $\gamma=\widehat\gamma=0.99$ and $N_p=2$.
}
\label{simhbt}
\end{center}
\end{figure}

\subsubsection{EBCM simulation}

In Fig.~\ref{simhbt} we present the EBCM simulation results of the HBT experiment depicted in Fig.~\ref{exphbt}
with messengers that contain the time-of-flight and the angle $\xi$, see Eq.~(\ref{mess2}).
For simplicity, we have put detector $D_1$ at $(X,0)$ and plot the single detectors and
coincidence counts as a function of the $y$-position of detector $D_0$.
In each simulation step, both sources $S_0$ and $S_1$ create a messenger 
with their initial time-of-flight set to some randomly chosen messages $\mathbf{y}_{n}$ ($n=0,1$)
which are kept fixed for $N_{\mathrm{F}}$ successive pairs of messengers.
Two pseudo-random numbers are used to determine whether the messengers travel to detector $D_0$ or $D_1$.
The time-of-flights are given by
\begin{equation}
T_{n,m}=\frac{\sqrt{X^2+((1-2n)d/2 - y_m)^2}}{c}
,
\label{hbt4a}
\end{equation}
where $n=0,1$ and $m=0,1$ label the source and detector, respectively.
As Fig.~\ref{simhbt} shows, averaging over the randomness in the initial messages wipes
out all interference fringes in the single-detector counts, in agreement with \MT.
Denoting the total number of emitted pairs of messengers by $N_{\mathrm{tot}}$,
we find that the number of single-detector counts fluctuates around $N_{\mathrm{tot}}/2$,
as expected from \MT, providing numerical evidence that the detectors have indeed 100\% detection efficiency.
Similarly, the data for the coincidence counts are in excellent agreement with the expression
\begin{equation}
N_{\mathbf{coincidence}} = \frac{N_{\mathrm{tot}}}{8} \left(1+ \frac{1}{2}\cos 2\pi f \Delta T\right)
,
\label{hbt4}
\end{equation}
predicted by \MT.

For simplicity, we have confined the above presentation to the case of a definite polarization.
Simulations with randomly varying polarization (results not shown) are also in concert
with \MT.
Elsewhere, we show that three-photon interference can be modeled by an EBCM as well~\cite{JIN10a}.

\subsubsection{Nonclassicality}

From Eq.~(\ref{hbt4}), it follows that the visibility of the interference fringes, defined by
\begin{equation}
{\cal V}= \frac{\max(N_{\mathbf{coincidence}})-\min(N_{\mathbf{coincidence}})}{\max(N_{\mathbf{coincidence}})+\min(N_{\mathbf{coincidence}})}
,
\label{hbt5}
\end{equation}
cannot exceed 50\%.
It seems commonly accepted that the visibility of a two-photon interference experiment
exceeding 50\%
is a criterion of the nonclassical nature of light.
On the other hand, according to Ref.~\cite{AGAF08}, the existence of high-visibility interference
in the third and higher orders in the intensity cannot be considered as a signature of three- or four-photon interference.
Thus, it seems that the two-photon case may be somewhat special, although there is no solid argument why this should be so.

As in the case of the EPRB experiment, HBT experiments employ time-coincidence to measure
the intensity-intensity correlations.
It is therefore quite natural to expect that a model that purports to explain
the observations accounts for the time delay
that occurs between the time of arrival at a detector and the actual click of that detector.
In  \QT, time is not an observable and can therefore not be computed within the theory proper,
hence there is no way that these time delays, which are being measured, can be accounted for by  \QT.
Consequently, any phenomenon that depends on these time delays must find an explanation
outside the realm of  \QT\ (as it is formulated to date).

It is straightforward to add a time-delay mechanism to the EBCM of the detector.
As an illustrative example, we assume that the time of the detector click,
relative to the time of emission, is given by
\begin{equation}
t_{\mathrm{delay}}=T_{n,m}+rT_{\mathrm{max}}(1-|\mathbf{T}|^2)^h
,
\label{hbt6}
\end{equation}
where $0<r<1$ is a pseudo-random number, and $\mathbf{T}$ is given by Eq.~(\ref{trans2}).
The time scale $T_{\mathrm{max}}$ and the exponent $h$ are
free parameters of the time-delay model.
Coincidences are counted by comparing the difference between the delay times
of detectors $D_0$ and $D_1$ with a time window $W$, exactly in the same way
as is done in the EPRB experiment (see Eq.~(\ref{eprb2}).

\begin{figure}[t]
\begin{center}
\includegraphics[width=8cm]{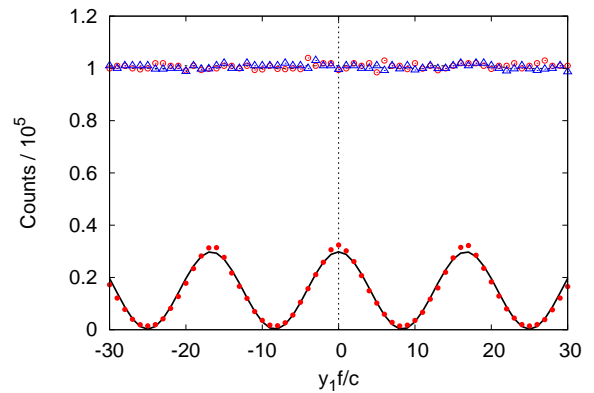}
\caption{%
Computer simulation data of the single-particle and two-particle counts
for the HBT experiment depicted in Fig.~\ref{exphbt}, generated
by the same EBCM that produced the data of Fig.~\ref{simhbt}
extended with the time-delay model Eq.~(\ref{hbt6}).
Red open circles: EBCM results for the counts of detector $D_0$.
Blue open triangles: EBCM results for the counts of detector $D_1$.
Red closed circles: EBCM results for the coincidence counts.
The dashed and solid lines are least-square fits of the predictions of wave theory
to the EBCM data for the single detector and coincidence counts, respectively.
Simulation parameters: $N_{\mathrm{tot}}=2\times 10^5$ events per $y_1f/c$-value,
$N_{\mathrm{F}}=40$, $X=100000c/f$, $d=2000c/f$, $\gamma=\widehat\gamma=0.99$, $N_p=2$,
$W=2/f$, $T_{\mathrm{max}}=2000/f$ and $h=8$.
}
\label{simhbt2}
\end{center}
\end{figure}

From the simulation results presented in Fig.~\ref{simhbt2},
it is clear that by taking into account that there are fluctuations
in the time delay that depend on the time-of-flight and the internal state
of the detector, the visibility changes from ${\cal V}=50\%$ to ${\cal V}\approx100\%$.
In Fig.~\ref{simhbt2}, the solid line represents the least square fit
of $ a(1+b\cos2\pi f\Delta t) $ to the simulation data.

This example demonstrates that a purely classical corpuscular model of a two-photon interference experiment
can yield visibilities that are close to 100\%,
completing the picture that high visibilities in two-, three- or four-photon interference experiments can be explained within
the realm of a classical theory, such as an EBCM.
In any case, the commonly accepted criterion of the nonclassical nature of light needs to be be revised.

The time delay model Eq.~(\ref{hbt6}) is perhaps one of the simplest that yield interesting results but
it is by no means unique. The study of various time delay models with applications
to fermion and boson statistics will be published elsewhere.

\section{Summary and outlook}\label{summary}

We have demonstrated that one universal event-based corpuscular model (EBCM) for the interaction of photons
with matter suffices to explain the interference and correlation
phenomena that are observed when individual photons are detected one by one.
Of course, this model produces the frequency distributions for observing many photons
that are in full agreement with the predictions of \MT\ and  \QT.
This is not surprising: \MT\ is used as a guiding principle to determine the single-event dynamics of the model.
The main conclusion of these simulations is that the effects such as EPR correlations and enhanced visibility
in two-photon interference in the HBT experiment are contained in a corpuscular model of light
and simply appear as the result of the particular processing of the detection events.
The EBCM is entirely classical in the sense that
it uses concepts of the macroscopic world and makes no reference to  \QT\
but is nonclassical in the sense that it does not rely on the rules of classical Newtonian dynamics.

It is remarkable that the observations of all the fundamental quantum optics experiments
covered in this paper can be explained using a computer simulation model that is elementary.
All that is required is elementary algebra and some Euclidian geometry.
This is in sharp contrast with the mathematical machinery that has to be mastered in order
to understand the meaning of the symbols used to formulate \MT\ or  \QT.

One may agrue that a finite-difference time-domain (FDTD) Maxwell solver~\cite{TAFL05} also carries out
very simple arithmetic operations but the point is that the simple update rules
for the electromagnetic-field variables are derived from a set of differential (or integral) equations.
In contrast, in the EBCM approach, the rules are constructed such that
the stationary state of the classical, dynamical system of processing
units yields the intensity distribution of the wave problem, not the other way around.
In other words, for quantum optics experiments, \MT\ and \QT\ arecontained in the EBCM
presented in this paper. We consider it unlikely that this EBCM is unique:
There may be several ways to modify the update rule of the processors such
that the stationary state of the EBCM agrees with the prediction of \MT,

An important question is whether the event-based corpuscular model can make predictions
that can be tested experimentally.
From the nine examples presented in this work, it is clear that
after a few hunderds of photons have been processed by the EBCM,
the frequencies of observations are hardly distinghuisable from
the intensities expected from \MT\ (recall that the EBCM of the detector has 100\% detection efficiency)
Therefore, to test the EBCM, one has to concieve an experiment that
is capable of testing the transient regime of the EBCM,
that is the regime before the EBCM reaches its stationary state,.
Elsewhere we have proposed an experiment with a Mach-Zehnder interferometer that
might be used for this purpose~\cite{MARC10}.
We hope that our simulation results will stimulate the design of new time-resolved single-photon
experiments to test our corpuscular model for optical phenomena.

We would like to draw attention to the fact that the
EBCM approach may have practical applications as a new method
for simulating optical phenomena.
From a computation viewpoint, the salient features of our event-based approach are:

\begin{itemize}
\item{It yields the stationary solution of the Maxwell equations
by simulating particle trajectories only.}
\item{Material objects are represented by DLM-based units placed on a boundary
of these objects, which in practice involves some form of discretization. Disregrading this discretization,
all calculations are performed using Euclidean geometry.
}
\item{Artifacts due to the unavoidable termination of the simulation volume which are inherent
to wave equation solvers~\cite{TAFL05} are absent: Particles that
leave the simulation volume can simply be removed from the simulation.}
\item{Unlike wave equation solvers which may consume substantial computational resources (i.e. memory and CPU time)
to simulate the propagation of waves in free space, it calculates the motion of the corresponding particles in free space
at almost no computational cost.}
\item{Modularity: Starting from the unit that simulates the behavior of a
plane interface between two homogeneous media other optical components
can be constructed by repeated use of the same unit.}
\end{itemize}

The work presented in this paper may open a route to
rigorously include the effects of interference in ray-tracing software.
For this purpose, it may be necessary to extend the DLM-based model
for a lossless dielectric materials to, say a Lorentz model for the response of material
to the electromagnetic field~\cite{TAFL05} and to extend the model to simulate phenomena that,
in \MT, are due to evanescent waves.
We leave these extensions for future research.

\section{Aknowledgement}
We would like to thank K. De Raedt and S. Miyashita for many helpful comments.
This work is partially supported by NCF, the Netherlands.

\appendix
\section*{Deterministic Learning Machines}\label{dlms}
We begin by asking the following simple question:
Given a real number $0\le y \le 1$ how can we
generate a sequence of binary variables
$y_1,\ldots,y_K=0,1$ such that
their average $(y_1+\ldots+y_K)/K$ yields a good (in a certain sense to be defined later)
approximation for $y$?

\subsection{Pseudo-random machine}\label{DLMA}

Perhaps the most obvious answer to this question is the following:
Use a pseudo-random number generator to
produce numbers $0< r_1,\ldots,r_K \le 1$
and to set $y_k=0$ ($y_k=1$) if $r_k\le y$ ($r_k> y$).
If the pseudo-random number generator is any good,
we expect that $(y_1+\ldots+y_K)/K\approx y$
if $K$ is large but much less than the period
of the pseudo-random number generator~\cite{PRES03}.
Theoretically, we can analyze the properties of these sequences
by imagining that there exist a device that generates random variables in the Kolmogorov sense.
Then, we would have
\begin{equation}
\lim_{K\rightarrow\infty}\frac{1}{K}\sum_{k=1}^K y_k\stackrel{P}{\rightarrow}y
,
\label{dlms0}
\end{equation}
where the notation $A\stackrel{P}{\rightarrow} B$ means that
$A$ is equal to $B$ with probability one,
which in plain words means it is likely that $A=B$
but that it can happen that $A\not=B$.
For sufficiently large $K$, probability theory
tells us that with $K$ binary variables, we can represent of the order of $1/c(d)\sqrt{K}$
real numbers, where $c(d)$ is proportional to
the number of significant digits $d$ of these real numbers~\cite{RAED06a}.

Thus, when we use (pseudo) random numbers, the problem
can be solved in a rather trivial manner.
But can we do better and dispose of random numbers altogether?
Yes, we can and perhaps surprisingly, the algorithm~\cite{RAED05b} that follows
is not more complicated than the simplest algorithms
that are used to generate pseudo random numbers~\cite{PRES03}.

\subsection{Deterministic learning machine I}\label{DLMB}

Let us consider a machine that receives as input messages in the form
of numbers $0\le y_k\le 1$ for $k=1,2,\ldots$.
For each $y_k$ that it receives, the machine updates its internal state,
represented by $0\le x_k \le 1$ according to
\begin{equation}
x_{k}=\gamma x_{k-1}+(1-\gamma )\Delta_{k},
\label{dlms1}
\end{equation}
where $\Delta_k = \Theta(|\gamma x_{k-1}+(1-\gamma )-y_k|-|\gamma x_{k-1}-y_k|)$
with $\Theta(.)$ denoting the unit step function.
In words, the machine will choose the new internal state $x_{k}$
such that the difference with the input $y_k$ is minimum~\cite{RAED05b}.
The sequence $\Delta_1,\Delta_2,\ldots$ are the output messages of the machine.
The parameter $0\le \gamma<1$ controls both the speed
and accuracy with which the machine can learn
the input value $y_k$~\cite{RAED05b,RAED06a}.
As a detailed mathematical analysis of the dynamics of the machine
defined by the rule Eq.~(\ref{dlms1}) is given in Ref.~\onlinecite{RAED06a},
it will not be repeated here.

\begin{figure}[t]
\begin{center}
\includegraphics[width=8cm]{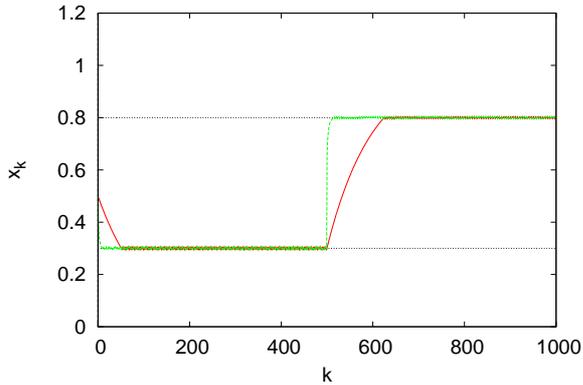}
\caption{%
The internal state of the DLM as a function of the number of input events $k$.
Red line: DLM defined by the update rule Eq.~(\ref{dlms1}).
Green line: DLM defined by the update rule Eq.~(\ref{dlms2}).
}%
\label{figdlm2}
\end{center}
\end{figure}

Here, we illustrate the operation of the DLM defined by update rule
Eq.~(\ref{dlms1}) by a simple example in which we feed
the DLM by a sequence of 500 messages with fixed value $y_k=0.3$
followed by another 500 messages with fixed value $y_k=0.8$.
Initially the machine is in the state $x_0=1/2$ and $\gamma=0.99$.

In Fig.~\ref{figdlm2} (red line) we show
simulation results of the internal state $x_k$ of the DLM as
a function of the number of events $k$ that have been received by the DLM.
From Eq.~(\ref{dlms1}), $x_0=0.5$ and $y=0.3$
it immediately follows that initially $\Delta_k=0$
and that it takes about $k_0=(\log 0.3 -\log 0.5)/\log 0.99\approx51$
events for the internal vector to reach a state where $x_k\approx y_k=0.3$
(for $k\le500$), in concert with the data of Fig.~\ref{figdlm2} (red line).

Once the DLM has reached this regime, its behavior changes drastically:
With $y_k$ still fixed, its internal vector will start oscillating about $y_k$.
If $x_k$ decreases (increases), the machine generates $\Delta_k=0(1)$ as output~\cite{RAED05b,RAED06a}.
It can be shown that in this stationary regime (and this regime only)
with fixed input value $y_k=y$, the average $K^{-1}\sum_{k=1}^K\Delta_k = N/K$
is an optimal, rational approximation to $y$, optimal in the sense that the variance
of the bit stream is minimal, implying that subsequences
of the $\Delta_k$'s yield obviously less accurate but again optimal approximations~\cite{RAED05b,RAED06a}.
In plain words, the sequence of $\Delta_k$'s is such that each subsequence yields an approximation to
the fixed input $y_k=y$.

From Fig.~\ref{figdlm2} (red line), it is clear that if we change the input from $y_k=0.3$ for $k\le 500$
to $y_k=0.8$ for $501\le k \le 1000$,
the machine needs about eighty input events to reach the new stationary state
This relaxation process would take even much longer if we would have chosen $\gamma=0.999$ instead of $\gamma=0.99$.
Thus, for this very simple machine, there is a trade off between accuracy and the time it needs
to respond to significant changes of the input value.

It is not difficult to modify the machine such that it responds much faster
to changes of the input value.
As one particular example of such a modification, consider the machine defined by the rules
\begin{eqnarray}
\gamma'&=&\min(1-|x_{k-1}-y_k|,\gamma),
\label{dlms2a}
\\
x_{k}&=&\gamma' x_{k-1}+(1-\gamma')\Delta_{k}.
\label{dlms2}
\end{eqnarray}
In Fig.~\ref{figdlm2} (green line), we present the results of a simulation in which we feed
the DLM defined by Eqs.~(\ref{dlms2a}) and (\ref{dlms2}) by a sequence of 500 messages with fixed value $y_k=0.3$
followed by another 500 messages with fixed value $y_k=0.8$.
Comparing the results of the DLMs defined by Eq.~(\ref{dlms1}) and Eqs.~(\ref{dlms2a}),(\ref{dlms2}),
respectively, it is clear that the former responds much slower to
changes in the input than the latter.

It should be noted that there is nothing special about the DLM defined by Eq.~(\ref{dlms2a}):
There are many ways to improve the response time by letting $\gamma'$ depend on the input, the current state and
$\gamma$, some examples being given later.
It should be noted though that if DLMs are used
to construct event-based processes that reproduce the results of wave mechanics,
the only requirement is that these results are being reproduced
in the regime where the DLMs have reached the stationary state, that is
after many events have been processed.

\subsection{Deterministic learning machine II}\label{DLMC}

The DLM of Section~\ref{DLMA} encodes the data (a real number) contained in messages
it receives in its internal state and for each message it receives, generates as output a one-bit message.
If the input messages are constant over a sufficiently long period,
the stream of output bits is an accurate representation of the input data.
From the description of this machine, it is rather obvious that it can be generalized
to accept messages that contain several real numbers.

In this section, we consider a different type of DLM, namely a DLM that updates its
internal state such that it represents the average of the data contained
in the input messages.
Specifically, we consider messages containing as data $d$-dimensional unit vectors $\mathbf{v}_k$
(Euclidian norm $\Vert \mathbf{v}_{k} \Vert=1$)
and a DLM with an internal state represented by a $d$-dimensional vector $\mathbf{x}_{k}$.

The state of the DLM is updated according to the rule
\begin{equation}
\mathbf{x}_{k}=\gamma \mathbf{x}_{k-1}+( 1-\gamma ) \mathbf{v}_k
,
\label{dlms3}
\end{equation}
where $0<\gamma <1$ and Euclidian norm of
the initial internal state is assumed to satisfy $\Vert \mathbf{x}_{0} \Vert\le1$.
The formal solution of Eq.~(\ref{dlms3}) reads,
\begin{equation}
\mathbf{x}_{k}=\gamma^{k-k'} \mathbf{x}_{k'}+( 1-\gamma ) \sum_{j=k'+1}^{k}\gamma^{k-j}\mathbf{v}_j
,
\label{dlms4}
\end{equation}
for all $0\le k'<k$.
From Eq.~(\ref{dlms4}) it immediately follows that
$\Vert \mathbf{x}_{k} \Vert\le\gamma^k\Vert \mathbf{x}_{0} \Vert+1-\gamma^k\le1$
for all $k\ge 1$, meaning that the vectors representing the internal state of the DLM
can never leave the $d$-dimensional sphere of radius one.
In other words, the update rule Eq.~(\ref{dlms4}) defines a numerically stable
iterative procedure.

Note the amount of internal storage that this DLM uses
is ``as large'' as the storage needed to represent the data contained in a single message
and that the last term Eq.~(\ref{dlms4}) has the structure of a convolution
of the input data $\mathbf{v}_j$ and a ``memory'' kernel $\gamma^{j}$.

From the formal solution Eq.~(\ref{dlms4}),
the fact that in practice the sequence $\{{\bf v}_1, {\bf v}_2, \cdots, {\bf v}_K \}$ is finite,
and the usual trick to assume a periodic continuation of the sequence, we have
\begin{eqnarray}
{\bf x}_{mK}&=&\gamma^K {\bf x}_{(m-1)K}+( 1-\gamma )\sum_{j=(m-1)K+1}^{mK}\gamma^{mK-j}{\bf v}_{j}
\nonumber \\
&=&\gamma^K {\bf x}_{(m-1)K}+( 1-\gamma )\sum_{j=1}^{K}\gamma^{K-j}{\bf v}_{j+(m-1)K}
\nonumber \\
&=&\gamma^K {\bf x}_{(m-1)K}+( 1-\gamma ){\bf f}_K
,
\label{dlms5}
\end{eqnarray}
where
\begin{equation}
{\bf f}_K=\sum_{j=1}^{K}\gamma^{K-j}{\bf v}_{j},
\end{equation}
and $m\geq0$. From Eq.~(\ref{dlms5}) we find
\begin{equation}
{\bf x}_{mK}=\gamma^{mK} {\bf x}_0+( 1-\gamma )\frac{1-\gamma^{mK}}{1-\gamma^K}{\bf f}_K,
\end{equation}
and hence
\begin{equation}
\lim_{m\rightarrow\infty}{\bf x}_{mK}=\frac{1-\gamma}{1-\gamma^K}\sum_{j=1}^{K}\gamma^{K-j}{\bf v}_{j},
\end{equation}
such that
\begin{equation}
\lim_{\gamma\rightarrow 1^-}\lim_{m\rightarrow\infty}{\bf x}_{mK}=\frac{1}{K}\sum_{j=1}^{K}{\bf v}_{j}.
\label{dlms6}
\end{equation}
From Eq.~(\ref{dlms6}), we conclude that as $\gamma\rightarrow 1^-$
the internal vector converges to the time average of the vectors
$\mathbf{v}_1, \mathbf{v}_2, \cdots, \mathbf{v}_K$.

Some analytical insight into the behavior of this DLM can be obtained
by assuming that $\mathbf{x}_k$ and $\mathbf{v}_k$ are
the values of time-dependent vectors $\mathbf{x}(t)$ and $\mathbf{v}(t)$ sampled at regular
time intervals $\tau$.
If $\mathbf{x}(t)$ allows a Taylor series expansion, we may write
$\mathbf{x}_k=\mathbf{x}(\tau k)$,
$\mathbf{x}_{k-1}=\mathbf{x}(\tau k)-\tau d\mathbf{x}(t)/dt|_{t=\tau k}+{\cal O}(\tau^2)$ such
that the update rule Eq.~(\ref{dlms3}) can be expressed as
\begin{eqnarray}
\frac{d \mathbf{x}(t)}{dt}=-\frac{1-\gamma}{\tau\gamma}\mathbf{x}(t)+\frac{1-\gamma}{\tau\gamma}\mathbf{v}(t)
.
\label{dlms7}
\end{eqnarray}
In order that Eq.~(\ref{dlms7}) makes sense for $\tau\rightarrow0$, we must have
$\lim_{\tau\rightarrow0}(1-\gamma)/\tau\gamma=\Gamma$.
This requirement is trivially satified by putting $\gamma=1/(1+\tau\Gamma)$.
Then Eq.~(\ref{dlms7}) takes the form of the first-order linear differential equation
\begin{eqnarray}
\frac{d \mathbf{x}(t)}{dt}=-\Gamma\mathbf{x}(t)+\Gamma\mathbf{v}(t)
.
\label{dlms8}
\end{eqnarray}
The formal solution of Eq.~(\ref{dlms8}) reads
\begin{eqnarray}
\mathbf{x}(t)=e^{-t\Gamma}\mathbf{x}(0)+\Gamma\int_0^t e^{-u\Gamma}\mathbf{v}(t-u)du
.
\label{dlms9}
\end{eqnarray}
As in Eq.~(\ref{dlms4}), the last term Eq.~(\ref{dlms9}) has the structure of a convolution
of the input data $\mathbf{v}(u)$ and the memory kernel $e^{-u\Gamma}$.
From the derivation of Eq.~(\ref{dlms8}), it follows that
if we interpret $\tau$ as the time interval between two successive messages
and let $\tau$ approach zero, then $\gamma=1/(1+\tau\Gamma)$ approaches one and the DLM defined by the update rule
Eq.~(\ref{dlms3}) ``solves'' the differential equation Eq.~(\ref{dlms8}).
Therefore, we may view Eq.~(\ref{dlms8}) as a course-grained, continuum approximation to the discrete,
event-by-event process defined by Eq.~(\ref{dlms3}).

\begin{figure}[t]
\begin{center}
\includegraphics[width=8cm]{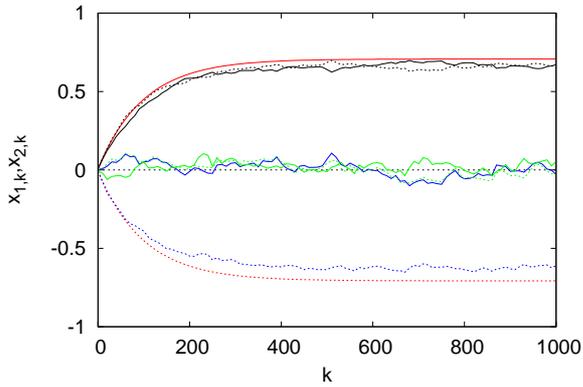}
\caption{The two components of the internal vector ${\mathbf x}_{k}=(x_{1,k},x_{2,k})$
as a function of the number of received events $k$ for four different input messages
${\mathbf e}_{k}=(1,-1)/\sqrt{2}$ (red lines) with statistical average $(1,-1)/\sqrt{2}$,
${\mathbf e}_{k}=(r^{1/2}_k,(1-r_k)^{1/2})$ (black lines) with statistical average $(2/3,2/3)$,
${\mathbf e}_{k}=(\cos\pi r_k,-\sin\pi r_k)$ (blue lines) with statistical average $(0,2/3)$,
and ${\mathbf e}_{k}=(\cos2\pi r_k,\sin2\pi r_k)$ (green lines) with statistical average $(0,0)$,
$0\le r_k<1$ being uniform pseudo-random numbers.
Solid lines: $x_{1,k}$; Dashed lines: $x_{2,k}$.
In all cases ${\mathbf x}_{0}=(0,0)$ and $\gamma=0.99$.
}
\label{figdlm3}
\end{center}
\end{figure}

\begin{figure}[t]
\begin{center}
\includegraphics[width=8cm]{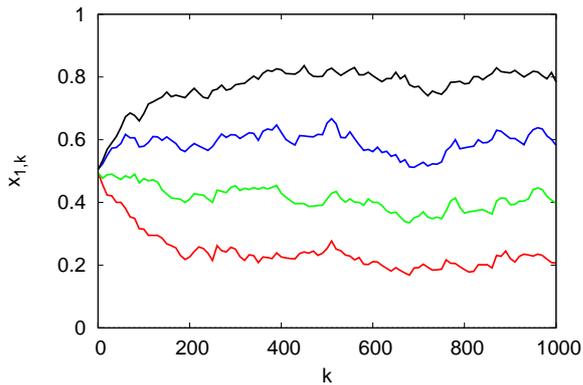}
\caption{The first components of the internal vector ${\mathbf x}_{k}=(x_{1,k},x_{2,k})$
of a DLM that has two input ports to receive messages.
A message arriving on input port 0(1) is represented by $\mathbf{v}_k=(1,0)$ ($\mathbf{v}_k=(0,1)$).
The probability that a message arrives on input port 0(1) is $p_0$ ($1-p_0$).
Initially, the DLM is in the state ${\mathbf x}_{0}=(1/2,1/2)$ and $\gamma=0.99$.
Red line: $p_0=0.2$;
Green line: $p_0=0.4$;
Blue line: $p_0=0.6$;
Black line: $p_0=0.8$.
This DLM estimates the frequency with which the messages arrive on input ports 0 and 1.
}
\label{figdlm4}
\end{center}
\end{figure}

As an illustration of how this DLM can be used, we consider two simple cases:
\begin{enumerate}
\item{%
A DLM with one input port receiving messages represented by a two-dimensional unit vector
$\mathbf{v}_k$.
According to Eq.~(\ref{dlms6}), after the DLM has received enough events such that it
reached its stationary state, its internal state represents the time average
of the events, that is we have ${\bf x}_K\approx K^{-1}\sum_{k=1}^K \mathbf{v}_k$
of $\gamma\lesssim 1$.
In Fig.~\ref{figdlm3}, we present some simulation results for different random
vectors $\mathbf{v}_k$. In all cases, the internal state shows the expected
behavior, namely it converges to the time-average of the input data.
Note that the DLM has no information about the total number of events.
}%
\item{%
A DLM that receives messages (the precise content of which is of no importance here)
on either input port ``0'' or input port ``1'' (but never on both ports simultaneously)
and has its initial state set such that $x_{1,0}+x_{2,0}=1$.
Then, setting $\mathbf{v}_k=(1,0)$ ($\mathbf{v}_k=(0,1)$)
if the message arrives thought port 0 (1),
it follows directly from Eq.~(\ref{dlms6}) that
$x_{1,k}+x_{2,k}=1$ for all $k\ge1$ and that as a result, the
internal state of the DLM will converge to the relative frequencies
with which the messages arrive on ports 0 and 1.
Some illustative simulation results are presented in Fig.~\ref{figdlm4},
showing that this DLM can be employed to estimate from a time series
of two different types of events, the relative frequency of these events.
}%
\end{enumerate}

For later applications, it is of the utmost importance that
these DLMs can estimate averages without having
to count the total number of events (which in real optics experiments is not known).
From Figs.~\ref{figdlm2} and \ref{figdlm3}, it is clear
that it may take quite a number of event for the DLMs
to reach their stationary state.
If necessary, the relaxation time can be reduced (significantly)
by adding a rule that changes $\gamma$, analogous to the trick
we introduced in Section~\ref{DLMB}~\cite{JIN09a}.

\bibliography{/d/papers/epr}

\end{document}